%% file: main.tex
\documentclass[10pt,journal,compsoc]{IEEEtran}
\ifCLASSOPTIONcompsoc
  \usepackage[nocompress]{cite}
\else
  \usepackage{cite}
\fi
\ifCLASSINFOpdf
\else
\fi
\usepackage[utf8]{inputenc} 
\usepackage[T1]{fontenc}    

\usepackage{epsfig}
\usepackage{graphicx}
\usepackage{amsmath}
\usepackage{amssymb}
\usepackage{enumitem}

\usepackage{booktabs}  
\usepackage{tabularx}  
\usepackage{multirow}  
\usepackage{xcolor}    
\usepackage{colortbl}  
\usepackage{pifont}    

\usepackage{bbding}

\usepackage[pagebackref,breaklinks,colorlinks]{hyperref}
\usepackage[capitalize]{cleveref}
\crefname{section}{Sec.}{Secs.}
\crefname{table}{Table}{Tables}
\crefname{figure}{Fig.}{Figs.}

\usepackage{xcolor}
\usepackage{tcolorbox}

\usepackage[linesnumbered,algo2e,boxed]{algorithm2e}
\graphicspath{{figs/}}
\usepackage[figuresright]{rotating}
\usepackage{amsmath,amssymb,textcomp,subfigure,multirow,upgreek}
\usepackage{booktabs}
\usepackage{color,mdwlist}
\usepackage{tikz}
\usepackage[edges]{forest}

\usepackage[edges]{forest}
\definecolor{hidden-draw}{RGB}{0,0,0}
\definecolor{hidden-pink}{rgb}{0.98, 0.94, 0.75}
\definecolor{level0}{rgb}{0.67, 0.88, 0.69}
\definecolor{level1}{rgb}{0.98, 0.92, 0.84}
\definecolor{level2}{rgb}{0.8, 0.8, 1.0}
\definecolor{level3}{rgb}{1.0, 0.71, 0.76}

\makeatletter
\DeclareRobustCommand\onedot{\futurelet\@let@token\@onedot}
\def\@onedot{\ifx\@let@token.\else.\null\fi\xspace}

\makeatother

\usepackage{ragged2e}

\usepackage{review}

\usepackage{graphicx,fontawesome}
\graphicspath{{figs/}}

\usepackage{caption}
\usepackage{wrapfig}

\usepackage{enumitem}
\setitemize{noitemsep,topsep=0pt,parsep=0pt,partopsep=0pt}

\begin{document}
\title{Earth Science Foundation Models: From Perception to Reasoning and Discovery
}

\author{Xiangyu Zhao, Bo Liu, Yuehan Zhang, Zelin Song, Wanghan Xu, \\Feng Liu, Fengxiang Wang, Ben Fei, Fenghua Ling, Wangxu Wei, Wenlong Zhang, Xiao-Ming Wu

\IEEEcompsocitemizethanks{\IEEEcompsocthanksitem Xiangyu Zhao, Bo Liu, Yuehan Zhang, Zelin Song and Xiao-Ming Wu are with the Department of Data Science and Artificial Intelligence, The Hong Kong Polytechnic University.
\IEEEcompsocthanksitem Wanghan Xu, Feng Liu, Fengxiang Wang, Ben Fei, Fenghua Ling, Wangxu Wei and Wenlong Zhang are with Shanghai Artificial Intelligence Laboratory, Shanghai, China.}
\thanks{Corresponding Authors: Wenlong Zhang (zhangwenlong@pjlab.org.cn) and Xiao-Ming Wu (xiao-ming.wu@polyu.edu.hk).}
}

%
%

\markboth{}%
{Shell \MakeLowercase{\textit{et al.}}: Bare Demo of IEEEtran.cls for Computer Society Journals}
%



\IEEEtitleabstractindextext{%
\input{draft/000_abstract}

\begin{IEEEkeywords}
Multimodal Large Language Models, Earth Science, GeoScience, Foundation Models
\end{IEEEkeywords}}

\maketitle

\IEEEdisplaynontitleabstractindextext

%
\IEEEpeerreviewmaketitle


%
%
%
%


\input{draft/010_intro}

\input{draft/020_overview}
\input{draft/030_MLLM}

\input{draft/040_discussion}

\input{draft/070_conclusion}

\footnotesize
\bibliographystyle{IEEEtran}
\bibliography{IEEEabrv, references}

\end{document}

%% file: draft/000_abstract.tex
\justify

\begin{abstract}


Large foundation models (FMs) are transforming Earth science by integrating heterogeneous multimodal data, such as multi-platform imagery, gridded reanalysis data, diverse geophysical and geochemical observations, and domain-specific text, to support tasks ranging from basic perception to advanced scientific discovery. This paper provides a unified review of Earth science foundation models (Earth FMs) through two complementary dimensions: \emph{depth}, which traces the evolution of model capabilities from perception to multimodal reasoning and agentic scientific workflows, and \emph{breadth}, which summarizes their expanding applications across the atmosphere, hydrosphere, lithosphere, biosphere, anthroposphere, and cryosphere, as well as coupled Earth system processes. Using this framework, we review representative multimodal Earth foundation models and compile more than 200 datasets and benchmarks spanning diverse Earth science tasks and modalities. We further discuss key challenges in multimodal data heterogeneity, scientific reliability and continual updating, scalability and sustainability, and the transition from foundation models to agentic and embodied Earth intelligence, and outline future directions toward more integrated, trustworthy, and actionable AI Earth scientists. Overall, this paper offers a structured roadmap for understanding the development of Earth foundation models from both capability depth and application breadth. To support community efforts, we maintain a curated list of related resources at: \url{https://github.com/xiangyu-mm/Cool-EarthScience-Foundation-Models}.

\end{abstract}

%% file: draft/010_intro.tex
\section{Introduction}
\label{sec:intro}

\begin{figure*}[ht]
	\centering
	\includegraphics[width=1.0\textwidth]{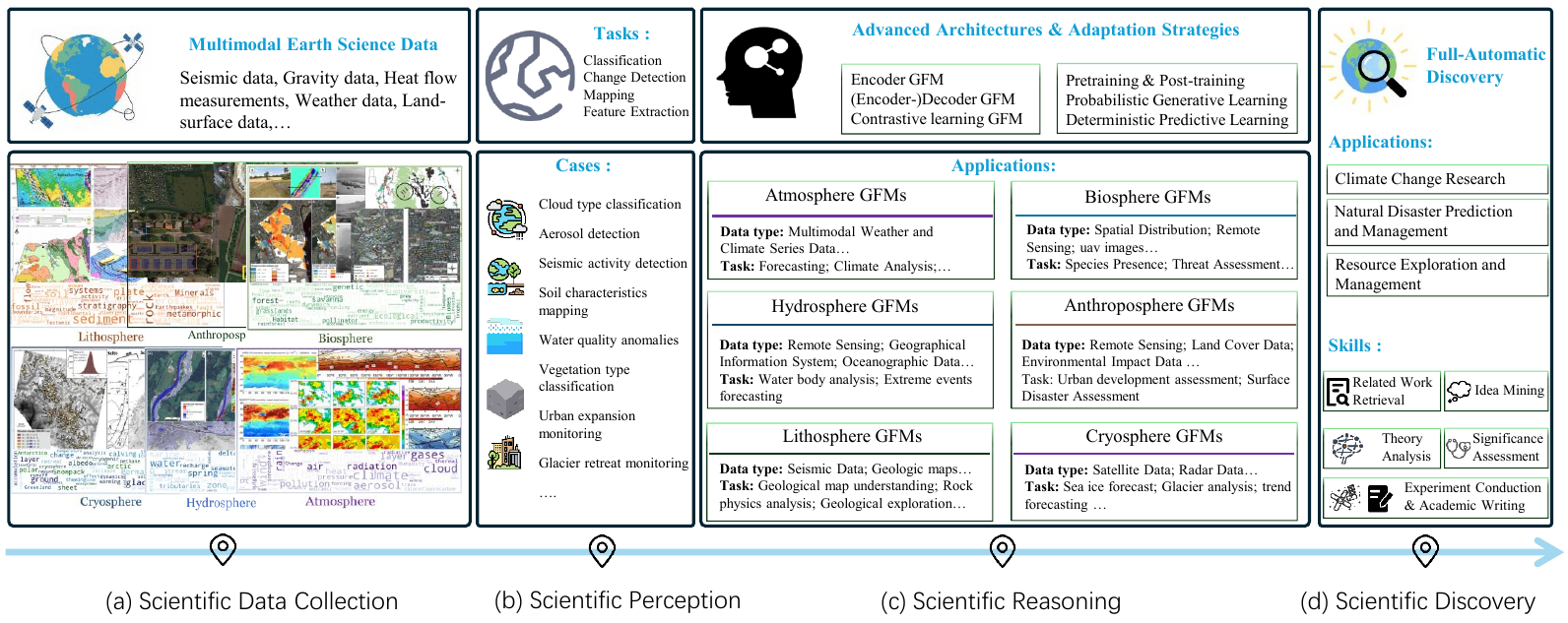}
        \caption{The evolutionary roadmap of Earth science AI. The field has progressed from multi-sphere \textit{Scientific Data Collection} and task-specific \textit{Scientific Perception} to advanced \textit{Scientific Reasoning} powered by Earth Science Foundation Models. Currently, it is transitioning toward fully automatic \textit{Scientific Discovery}, driven by AI agents capable of hypothesis generation, experiment execution, and autonomous analysis for complex environmental challenges.}
        \label{intro_case}
\end{figure*}

\IEEEPARstart{E}{arth} science has evolved from a set of largely independent subdisciplines into an integrated study of the coupled Earth system, including the atmosphere, hydrosphere, lithosphere, biosphere, anthroposphere, and cryosphere~\cite{zhao2024artificial}. As shown in Figure~\ref{intro_case}(a), the \textit{Scientific Data Collection} stage is characterized by the rapid growth of multimodal datasets, from seismic records to large-scale satellite imagery archives. These data create significant opportunities to improve understanding of environmental change, enhance hazard prediction, and support sustainability-oriented decision-making~\cite{bodnar2024aurora, bi2022pangu, guan2024citygpt}. However, their scale, heterogeneity, and cross-sphere complexity pose substantial challenges to conventional analytical workflows.

Reflecting the \textit{Scientific Perception} stage illustrated in Figure~\ref{intro_case}(b), early applications of artificial intelligence (AI) in Earth science~\cite{lam2023graphcast, mo2023s, haixin2023marinedet, zhang2021trs} were primarily developed for task-specific objectives, including land-cover classification, change detection, and feature extraction from remote sensing imagery. These approaches achieved notable success, particularly with the emergence of deep learning and vision-based models. Nevertheless, many core geoscientific problems extend beyond isolated perception tasks. They require the integration of heterogeneous observations across modalities and scales, reasoning over coupled physical processes, interaction with external scientific tools, and support for multi-step analytical workflows~\cite{shan2026climateagents,shabbir2026openearthagent}. Accordingly, AI in Earth science is shifting from narrowly specialized models toward more general and unified learning frameworks.

This transition has advanced the field into the era of \textit{Scientific Reasoning} (Figure~\ref{intro_case}(c)), stimulating growing interest in \emph{Earth science foundation models} (\textbf{Earth FMs}). These models are designed to learn transferable representations from large-scale geospatial, environmental, and scientific data spanning the major Earth spheres~\cite{schmude2024prithvi, jakubik2025terramind, bodnar2024aurora}. In contrast to conventional models tailored to a single dataset or modality, Earth FMs offer the potential to unify remote sensing imagery, geophysical fields, textual knowledge, and other Earth science data within a common modeling framework. Recent advances indicate that such architectures can support not only perception-oriented tasks, but also cross-modal understanding, complex scientific reasoning, and decision-relevant analysis.

Ultimately, the field is progressing toward fully automated \textit{Scientific Discovery}, as illustrated in Figure~\ref{intro_case}(d). Reasoning-oriented Earth AI, tool-augmented agents, and collaborative multi-agent systems are creating a path toward more executable and autonomous forms of scientific assistance~\cite{shabbir2026openearthagent, han2026gisclaw, shan2026climateagents, guo2025earthlink}. This trend aligns with the emerging vision of \emph{AI Earth Scientists}: intelligent systems capable of integrating heterogeneous data, analyzing scientific theories, conducting experiments, and supporting end-to-end scientific workflows. More broadly, these developments suggest a paradigm shift in which Earth AI is evolving from isolated predictive tools into integrated frameworks with the potential to address macro-scale challenges such as global climate change and natural disaster management autonomously.

Despite rapid progress, the literature on AI for Earth science remains fragmented and marked by two major limitations. First, existing surveys typically adopt a compartmentalized view of the Earth system, emphasizing isolated subdomains such as optical remote sensing or specific weather forecasting tasks~\cite{zhou2406towards, xiao2025foundation, zhu2026foundations}. This narrow perspective overlooks the intrinsically coupled nature of the Earth system, in which processes across different spheres interact dynamically. As a result, these surveys do not adequately capture the emerging shift toward unified, cross-sphere Earth foundation models~\cite{jakubik2025terramind, pantiukhin2026hierarchical, shabbir2026openearthagent}. Second, the current literature lacks a systematic analysis of the developmental trajectory of Earth FMs. Although several broader geoscientific reviews have been published, they focus primarily on \textit{Scientific Perception} and \textit{Reasoning}~\cite{zhao2024artificial, hong2026foundation}. A comprehensive roadmap is still lacking to characterize the evolution of these models from isolated predictive systems to dynamic, tool-augmented agents capable of autonomous \textit{Scientific Discovery}.


To address these gaps, this survey provides a holistic review of AI in Earth science along two complementary dimensions: application breadth and capability depth. Along the breadth dimension, we categorize existing Earth FMs across the six major Earth spheres—the atmosphere, hydrosphere, lithosphere, biosphere, anthroposphere, and cryosphere—as well as their coupled processes. Along the depth dimension, we trace the evolution of these models from basic perceptual capabilities to advanced multimodal reasoning and autonomous agentic workflows. Within this framework, we review representative models and compile more than 200 datasets spanning diverse modalities. Finally, we present a structured roadmap and discuss key challenges related to data heterogeneity, reliability, and scalability, with the aim of enabling integrated and actionable AI systems for Earth science.

Under this framework, the remainder of this survey is organized as follows. Section~\ref{sec:related_work} reviews the literature and positions this work relative to prior surveys, highlighting the transition toward Earth FMs. Section~\ref{sec:mllm_arch} introduces the necessary preliminaries, including the characteristics of diverse Earth science data and a three-stage view of Earth model evolution from scientific perception to reasoning and autonomous discovery. Section~\ref{sec:domain_fms} systematically reviews domain-specific Earth FMs across the six Earth spheres and summarizes representative datasets, tasks, and applications. Section~\ref{sec:extensions} examines the emerging paradigm of unified Earth FMs, with emphasis on multi-sphere integration, advanced reasoning, and tool-augmented collaborative agentic workflows. Section~\ref{sec:hallu} discusses key challenges, including data heterogeneity and physical consistency, and outlines future research directions. Finally, Section~\ref{sec:conclusion} concludes the survey.

\begin{table*}[htbp]
  \centering
  \caption{Comparison of Existing Surveys and Reviews with Our Work. \textit{Note:} \textbf{Perception}: Feature extraction, classification, static prediction. \textbf{Reasoning}: Logic inference, VQA, chain-of-thought. \textbf{Discovery}: Hypothesis generation, tool-use, agentic exploration. \textbf{RS}: Remote Sensing. \textbf{EO}: Earth Observation.}
  \label{tab:comparison}
  
  \renewcommand{\arraystretch}{1.4}
  \footnotesize
  
  \resizebox{\textwidth}{!}{%
  \begin{tabularx}{\textwidth}{@{} 
    l 
    >{\hsize=0.6\hsize\raggedright\arraybackslash}X 
    >{\hsize=0.6\hsize\raggedright\arraybackslash}X 
    >{\hsize=0.6\hsize\raggedright\arraybackslash}X 
    >{\hsize=2.2\hsize\raggedright\arraybackslash}X 
    @{}}
    \toprule
    \textbf{Ref. (Year)} & \textbf{Core Theme} & \textbf{Sphere Breadth} & \textbf{Capability Depth} & \textbf{Primary Focus \& Differentiation} \\
    \midrule
    
    Zhang et al. (2024)~\cite{zhang2024towards} & Urban FMs & Anthroposphere & Perception & Analyzes foundation models specifically tailored for smart city development and urban sensing. \\
    
    Zhao et al. (2024)~\cite{zhao2024artificial} & AI for Geoscience & Holistic & Perception \& Prediction & Comprehensively reviews AI across Earth spheres with a focus on physics-ML hybrids and prediction, but precedes the era of agentic workflows. \\
    
    Zhou et al. (2024)~\cite{zhou2406towards} & Vision-Language FMs & Geospatial (RS) & Perception & Investigates vision-language alignment methodologies applied to remote sensing imagery. \\
    
    Xiao et al. (2025)~\cite{xiao2025foundation} & RS Foundation Models & EO Data & Perception & Reviews pre-training architectures and downstream adaptation techniques for Earth observation tasks. \\
    
    Zhu et al. (2026)~\cite{zhu2026foundations} & Earth FMs & EO \& Climate & Perception \& Reasoning & Outlines the desired high-level attributes and theoretical frameworks for future Earth FMs. \\
    
    \midrule
    \rowcolor{gray!15} 
    \textbf{Ours (2026)} & \textbf{Earth FMs} & \textbf{Holistic} & \textbf{Perception $\to$ Discovery} & \textbf{Classifies the landscape according to Earth spheres and their corresponding modalities/task types, systematizing the shift toward autonomous agentic scientific discovery.} \\
    \bottomrule
  \end{tabularx}%
  }
\end{table*}

\input{figs/taxonomy}

%% file: figs/taxonomy.tex
\tikzstyle{my-box}=[
    rectangle,
    draw=hidden-draw,
    rounded corners,
    text opacity=1,
    minimum height=1.5em,
    minimum width=5em,
    inner sep=2pt,
    align=center,
    fill opacity=.4,
    line width=0.8pt,
]
\tikzstyle{leaf}=[my-box, minimum height=1.5em,
    fill=hidden-pink!80, text=black, align=left, font=\normalsize,
    inner xsep=2pt,
    inner ysep=4pt,
    line width=0.8pt,
]
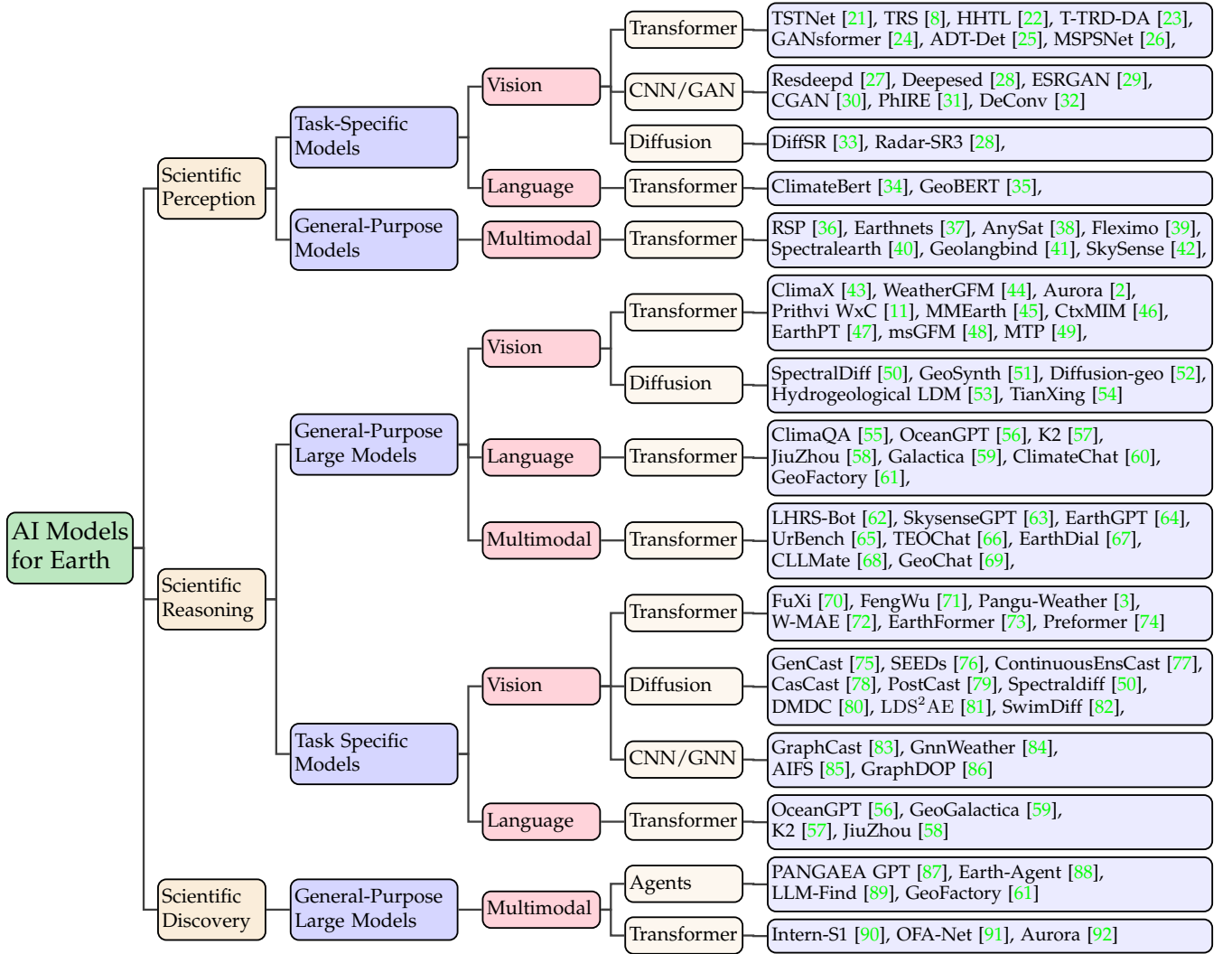
\begin{figure*}[t]
\centering
    \resizebox{0.98\textwidth}{!}{
        \begin{forest}
            forked edges,
            for tree={
                fill=level0!80,
                grow=east,
                reversed=true,
                anchor=base west,
                parent anchor=east,
                child anchor=west,
                base=left,
                font=\large,
                rectangle,
                draw=hidden-draw,
                rounded corners,
                align=left,
                minimum width=5em,
                edge+={darkgray, line width=1pt},
                s sep=3pt,
                inner xsep=2pt,
                inner ysep=3pt,
                line width=0.8pt,
                ver/.style={rotate=90, child anchor=north, parent anchor=south, anchor=center},
            },
            where level=1{text width=5em,font=\normalsize,fill=level1!90,}{},
            where level=2{text width=8em,font=\normalsize,fill=level2!80,}{},
            where level=3{text width=5.5em,font=\normalsize,fill=level3!60,}{},
            where level=4{text width=5.5em,font=\normalsize,fill=level1!40,}{},
            where level=5{text width=22em,font=\small,fill=level2!40,}{},
            [AI Models \\ for Earth
                [
                    Scientific \\ Perception
                    [
                        Task-Specific \\ Models
                        [
                            Vision
                            [
                                Transformer
                                [
                                    TSTNet~\cite{hao2022two}{,} 
                                    TRS~\cite{zhang2021trs}{,}
                                    HHTL~\cite{ma2022homo}{,} 
                                    T-TRD-DA~\cite{li2022transformer}{,} \\
                                    GANsformer~\cite{zhang2022gansformer}{,}
                                    ADT-Det~\cite{zheng2021adt}{,}
                                    MSPSNet~\cite{guo2021deep}{,}
                                ]
                            ]
                            [   
                                CNN/GAN
                                [
                                    Resdeepd~\cite{sharma2022resdeepd}{,}
                                    Deepesed~\cite{bano2022downscaling}{,}
                                    ESRGAN~\cite{watson2020investigating}{,}\\
                                    CGAN~\cite{wang2021fast}{,}
                                    PhIRE~\cite{stengel2020adversarial}{,}
                                    DeConv~\cite{hu2018deconvolution}
                                ]
                            ]
                            [   
                                Diffusion
                                [
                                    DiffSR~\cite{he2025diffsr}{,}
                                    Radar-SR3~\cite{bano2022downscaling}{,}
                                ]
                            ]
                        ]
                        [
                            Language
                            [   
                                Transformer
                                [
                                    ClimateBert~\cite{vaid2022towards}{,} 
                                    GeoBERT~\cite{liu2022few}{,}
                                ]
                            ]
                        ]
                    ]
                    [
                        General-Purpose \\ Models
                        [
                            Multimodal
                            [
                                Transformer
                                [
                                    RSP~\cite{wang2022empirical}{,} 
                                    Earthnets~\cite{xiong2022earthnets}{,}
                                    AnySat~\cite{astruc2025anysat}{,} 
                                    Fleximo~\cite{li2025fleximo}{,} \\
                                    Spectralearth~\cite{braham2025spectralearth}{,}
                                    Geolangbind~\cite{xiong2025geolangbind}{,}
                                    SkySense~\cite{guo2024skysense}{,}
                                ]
                            ]
                        ]
                    ]
                ]
                [
                    Scientific \\ Reasoning
                    [
                        General-Purpose \\ Large Models
                        [
                            Vision
                            [
                                Transformer
                                [
                                    ClimaX~\cite{nguyen2023climax}{,} 
                                    WeatherGFM~\cite{zhao2024weathergfm}{,}
                                    Aurora~\cite{bodnar2024aurora}{,} \\
                                    Prithvi WxC~\cite{schmude2024prithvi}{,} 
                                    MMEarth~\cite{nedungadi2024mmearth}{,}
                                    CtxMIM~\cite{zhang2023ctxmim}{,} \\
                                    EarthPT~\cite{smith2023earthpt}{,} 
                                    msGFM~\cite{han2024bridging}{,}
                                    MTP~\cite{wang2024mtp}{,}
                                ]
                            ]
                            [   
                                Diffusion
                                [
                                    SpectralDiff~\cite{chen2023spectraldiff}{,} 
                                    GeoSynth~\cite{sastry2024geosynth}{,}
                                    Diffusion-geo~\cite{cai2024diffusion}{,}\\
                                    Hydrogeological LDM~\cite{zhan2025toward}{,}
                                    TianXing~\cite{yuan2025tianxing}
                                ]
                            ]
                        ]
                        [
                            Language
                            [   
                                Transformer
                                [
                                    ClimaQA~\cite{manivannan2024climaqa}{,} OceanGPT~\cite{bi2023oceangpt}{,}
                                    K2~\cite{deng2024k2}{,} \\
                                    JiuZhou~\cite{chen2025jiuzhou}{,}
                                    Galactica~\cite{lin2023geogalactica}{,} 
                                    ClimateChat~\cite{chen2025climatechat}{,} \\
                                    GeoFactory~\cite{chen2025geofactory}{,}
                                ]
                            ]
                        ]
                        [
                            Multimodal
                            [   
                                Transformer
                                [
                                    LHRS-Bot~\cite{muhtar2024lhrs}{,} SkysenseGPT~\cite{luo2024skysensegpt}{,}
                                    EarthGPT~\cite{zhang2024earthgpt}{,} \\
                                    UrBench~\cite{zhou2025urbench}{,}
                                    TEOChat~\cite{irvin2024teochat}{,}
                                    EarthDial~\cite{soni2025earthdial}{,} \\
                                    CLLMate~\cite{li2024cllmate}{,}
                                    GeoChat~\cite{kuckreja2024geochat}{,}
                                ]
                            ]
                        ]
                    ]
                    [
                        Task Specific \\ Models
                        [
                            Vision
                            [
                                Transformer
                                [
                                    FuXi~\cite{chen2023fuxi}{,}
                                    FengWu~\cite{chen2023fengwu}{,}
                                    Pangu-Weather~\cite{bi2022pangu}{,} \\
                                    W-MAE~\cite{man2023w}{,}
                                    EarthFormer~\cite{gao2022earthformer}{,}
                                    Preformer~\cite{du2023preformer}
                                ]
                            ]
                            [   
                                Diffusion
                                [
                                    GenCast~\cite{price2025probabilistic}{,}
                                    SEEDs~\cite{li2024generative}{,}
                                    ContinuousEnsCast~\cite{andrae2024continuous}{,}\\
                                    CasCast~\cite{gong2024cascast}{,}
                                    PostCast~\cite{gong2024postcast}{,}
                                    Spectraldiff~\cite{chen2023spectraldiff}{,}\\
                                    DMDC~\cite{liu2022diffusion}{,}
                                    $\mathrm{LDS^{2}AE}$~\cite{qu2024lds2ae}{,}
                                    SwimDiff~\cite{tian2024swimdiff}{,}
                                ]
                            ]
                            [   
                                CNN/GNN
                                [
                                    GraphCast~\cite{lam2023learning}{,} GnnWeather~\cite{keisler2022forecasting}{,}\\
                                    AIFS~\cite{lang2024aifs}{,}
                                    GraphDOP~\cite{alexe2024graphdop}
                                ]
                            ]
                        ]
                        [
                            Language
                            [   
                                Transformer
                                [
                                    OceanGPT~\cite{bi2023oceangpt}{,} GeoGalactica~\cite{lin2023geogalactica}{,}\\
                                    K2~\cite{deng2024k2}{,}
                                    JiuZhou~\cite{chen2025jiuzhou}
                                ]
                            ]
                        ]
                    ]
                ]
                [   
                   Scientific \\ Discovery
                    [
                        General-Purpose \\ Large Models
                        [
                            Multimodal
                            [   
                                Agents
                                [
                                    PANGAEA GPT~\cite{pantiukhin2025accelerating}{,} 
                                    Earth-Agent~\cite{feng2025earth}{,}\\
                                    LLM-Find~\cite{ning2025autonomous}{,}
                                    GeoFactory~\cite{chen2025geofactory}
                                ]
                            ]
                            [
                                Transformer
                                [
                                    Intern-S1~\cite{bai2025intern}{,} 
                                    OFA-Net~\cite{xiong2024one}{,}
                                    Aurora~\cite{bodnar2025foundation}
                                ]
                            ]
                        ]
                    ]
                ]
            ]
        \end{forest}
    }
\caption{Taxonomy of AI models for Earth science across the three stages of \emph{Scientific Perception}, \emph{Scientific Reasoning}, and \emph{Scientific Discovery}, further organized by model scope, modality, and architecture family. Representative models are shown under each category.}
\vspace{-3mm}
\label{fig:taxonomy}
\end{figure*}

%% file: draft/030_MLLM.tex
\input{draft/031_data}

\input{draft/033_sphere_models}

%% file: draft/031_data.tex
\section{Related Surveys}
\label{sec:related_work}

The rapid evolution of AI in Earth science has led to a growing body of survey literature. Early reviews primarily cataloged deep learning applications for specific datasets, whereas more recent works have begun to recognize the emerging importance of foundation models. As summarized in Table~\ref{tab:comparison}, existing surveys generally fall into two broad categories: modality-centric reviews, predominantly focused on remote sensing, and broader reviews of AI paradigms in the geosciences. However, both categories exhibit important limitations in addressing the computational and systemic complexities of modern Earth FMs.

A substantial portion of the current literature treats Earth observation (EO) largely as an extension of traditional computer vision. Early surveys~\cite{jiao2023brain, zhang2023large} provided foundational overviews of pre-training methods for satellite imagery. More recent works, such as \cite{zhou2406towards} and \cite{xiao2025foundation}, have expanded this scope to include vision-language models and multimodal methods for EO. Although these surveys provide considerable technical depth on optical and non-optical imagery, they remain largely confined to perception-oriented tasks, such as classification, segmentation, and object detection. By focusing primarily on visual data, they underemphasize the cross-modal and cross-sphere representation learning needed to model Earth as an integrated system. They also give limited attention to highly heterogeneous non-visual data structures, such as spatiotemporal fields, irregular sensor networks, and in-situ observations, that are central to many Earth science domains.

Beyond modality-specific studies, broader reviews have examined the shift toward data-driven paradigms in the geosciences. For example, Zhao et al.~\cite{zhao2024artificial} discuss the transition from classical physics-based modeling to AI-driven methods and highlight their effectiveness in spatiotemporal prediction under data-scarce conditions. Similarly, recent perspective articles such as Zhu et al.~\cite{zhu2026foundations} have begun to articulate the desirable properties of an ideal Earth FM. However, these works are typically anchored either in conventional prediction settings or in high-level conceptual discussions, with limited attention to the computational pathways through which Earth FMs may evolve. In particular, they do not systematically examine the progression from perception and reasoning to tool-augmented, autonomous agents capable of orchestrating complex scientific workflows and supporting scientific discovery.

To address these gaps, this survey introduces a unified taxonomy of Earth AI along two complementary dimensions: \emph{application breadth} and \emph{capability depth}. Along the breadth dimension, we move beyond the conventional EO-centric view to cover computational challenges across \textbf{all six major Earth spheres}: atmosphere, hydrosphere, lithosphere, biosphere, cryosphere, and anthroposphere, while emphasizing cross-sphere multimodal integration. Along the depth dimension, we trace the evolution of Earth FMs from \textbf{perception} and \textbf{reasoning} to the emerging paradigm of agentic scientific \textbf{discovery}. By situating advanced LLM capabilities, including multi-step reasoning, tool use, and multi-agent coordination, within geophysical workflows, this survey aims to provide a comprehensive roadmap toward integrated and actionable AI systems for Earth science.

\section{Preliminaries}
\label{sec:mllm_arch}


\subsection{Earth Science Data}
Earth science spans multiple interconnected spheres, including the atmosphere, hydrosphere, biosphere, lithosphere, and anthroposphere. Consequently, the associated data resources exhibit extreme heterogeneity in modality, spatiotemporal scale, and semantic density. These variations stem from the distinct observational strategies and analytical paradigms historically adopted by isolated subdisciplines. To establish a unified framework for Earth-science-oriented foundation models, this section categorizes the primary data modalities: textual corpora, structured records, visual imagery, and time-series grids. As illustrated in Table~\ref{tab:dataset_all}, representative datasets vary substantially across Earth spheres not only in their sensing modalities and temporal coverage but also in their downstream computational tasks, ranging from deterministic forecasting to causal inference, semantic retrieval, and agentic question answering.

\textbf{Textual Resources.}
The growing use of large language models and reasoning agents in Earth science has made domain-specific textual corpora an important resource for both model development and evaluation. These materials support knowledge acquisition during pretraining, retrieval-based grounding, and the assessment of domain-specific reasoning. They span a broad range of formats, including textbooks, peer-reviewed literature, task-specific benchmarks, and real-time news or incident reports.

Textbooks in atmospheric science~\cite{mak2011atmospheric}, geology~\cite{crutzen2016geology}, oceanography~\cite{crust1977introduction}, and climate dynamics~\cite{change2001climate} provide structured and conceptually coherent coverage of core ideas, terminology, and causal relationships. As training resources, they can help models acquire foundational physical principles, standardized vocabulary, and common explanatory patterns. However, textbooks are inherently slow to reflect new discoveries, revised theories, and rapidly changing Earth-system conditions.

By contrast, peer-reviewed articles from venues such as \emph{Nature Geoscience} and the \emph{Journal of Geophysical Research} provide more current and specialized scientific knowledge. They contain recent methods, detailed uncertainty analyses, and quantitative findings across areas such as seismology, volcanology, and ecology. However, their technical density, including equations, tables, figures, and highly specialized terminology, makes them difficult to process reliably, especially in pipelines that require document parsing, retrieval, or text-figure alignment.

To support systematic evaluation, the community has developed task-specific textual benchmarks. In the climate domain, datasets such as CLIMATE-FEVER~\cite{diggelmann2020climate}, Climate-Stance~\cite{vaid2022towards}, ClimaText~\cite{varini2020climatext}, ClimateEval~\cite{kurfali2025climateeval}, and ClimaQA~\cite{manivannan2024climaqa} support quantitative evaluation on tasks such as claim verification, stance detection, and domain-specific question answering. Similarly, resources such as WaterER~\cite{xu2024unlocking} provide specialized entity-relation annotations for hydrology and water engineering. In addition, corpora derived from university examinations or professional geoscience certification tests can serve as compact, structured probes of factual recall and problem-solving ability.

Finally, real-time text streams, including news reports and emergency bulletins on hurricanes, floods, earthquakes, and droughts, provide timely context for disaster response, situational awareness, and event tracking. However, because these sources are noisy, incomplete, and uneven in scientific reliability, they require careful filtering, source validation, and temporal grounding. In retrieval-based or agent-oriented systems, using such streams effectively typically requires robust retrieval-augmented generation (RAG) pipelines that reconcile fast-changing reports with authoritative geoscientific sources. Overall, textual data are indispensable for endowing Earth-science models with domain knowledge and reasoning support, but their heterogeneous formats, uneven standardization, and trade-off between timeliness and scientific rigor remain significant challenges.

\textbf{Structured Data.}
Structured data provide an essential basis for knowledge organization, causal reasoning, and machine-readable representation in Earth science. Such data include relational databases, station records, event tables, knowledge graphs, and curated scientific repositories, many of which are closely linked to specific Earth spheres summarized in Table~\ref{tab:dataset_all}.

One representative direction is event-centered causal and relational modeling. CLLMate~\cite{li2024cllmate}, for example, aligns historical meteorological variables with weather and climate events by extracting entities, temporal information, locations, and impacts from large-scale news corpora and organizing them into a knowledge graph. This graph is then linked with gridded meteorological fields to support causal inference and event forecasting. Similar structured formulations are increasingly useful for connecting textual descriptions with numerical Earth-system data.

Station- and repository-based datasets are another major source of structured information. WEATHER-5K~\cite{han2024weather} organizes long-term observations from thousands of meteorological stations into a globally distributed network, enabling spatial-temporal analysis of atmospheric variability. Beyond meteorology, biodiversity repositories such as GBIF~\cite{telenius2011biodiversity} and ecological interaction databases such as GloBI~\cite{noori2026curated} support biosphere-oriented tasks including species distribution modeling and ecological network analysis. In the lithosphere, geochemical and petrological databases such as EarthChem~\cite{walker2005earthchem}, GEOROC~\cite{digis2021georoc}, and PetDB~\cite{lehnert2000global} provide structured records for studying crustal evolution, magmatic processes, and resource exploration. These examples illustrate that structured Earth science data are not limited to weather records, but span a broad range of disciplines and knowledge representations.

\textbf{Visual Resources.}
Visual data play a central role in Earth science because many geoscientific processes are inherently spatial and are often observed through imaging systems or scientific visualizations. These resources include remote sensing imagery, radar products, ecological maps, geological imaging, and multimodal image-text datasets.

Satellite remote sensing imagery is one of the most widely used visual data sources. Multispectral and hyperspectral observations from platforms such as MODIS~\cite{justice2002overview}, Landsat-9~\cite{masek2020landsat}, Sentinel-2~\cite{phiri2020sentinel}, ASTER~\cite{baldridge2009aster}, Hyperion~\cite{datt2003preprocessing}, and EMIT~\cite{green2023performance} capture diverse surface and atmospheric properties across visible, infrared, and microwave bands. These data support applications ranging from vegetation and wildfire monitoring to lithology mapping, mineral exploration, and hydrothermal alteration analysis. Radar-based products, including SAR imagery~\cite{moreira2013tutorial}, are especially valuable under cloud cover or low-visibility conditions, for example in flood mapping, terrain deformation analysis, and disaster monitoring.

Earth science visual resources also include event-centered and task-specific benchmarks. In atmospheric science, datasets such as SEVIR~\cite{veillette2020sevir}, MeteoNet~\cite{larvor2021meteonet}, and Digital Typhoon~\cite{kitamoto2023digital} support severe weather understanding and nowcasting. In hydrology, FloodNet~\cite{rahnemoonfar2021floodnet} provides UAV imagery paired with language annotations for post-flood scene understanding. In lithospheric applications, seismic imaging and tomographic datasets such as OpenFWI~\cite{deng2022openfwi}, OpenSWI~\cite{liu2025openswi}, and GlobalTomo~\cite{li2024globaltomo} offer visualized subsurface structures for interpretation and inversion. More recently, remote-sensing vision-language datasets such as RemoteCLIP~\cite{liu2024remoteclip}, RS5M~\cite{zhang2024rs5m}, RSIVQA~\cite{lobry2020rsvqa}, and SkyEye-968k~\cite{zhan2025skyeyegpt} have extended Earth science visual data toward multimodal pretraining, retrieval, captioning, and visual question answering. Despite their value, these data often require substantial preprocessing due to differences in spatial resolution, sensor characteristics, noise, and annotation quality.

\textbf{Time Series Resources.}
Time-series data describe the temporal evolution of Earth-system processes and are fundamental for understanding dynamics across the atmosphere, ocean, solid Earth, and ecosystems. They are typically sampled at regular or event-driven intervals and may be derived from direct observations, remote sensing systems, reanalyses, or physics-based data-assimilative models.

In atmospheric science, gridded reanalysis datasets such as ERA5~\cite{hersbach2020era5} and CRA5~\cite{han2024cra5} provide temporally continuous, spatially extensive reconstructions of variables including temperature, wind, pressure, and precipitation. These datasets are widely used for weather forecasting, climate diagnostics, and extreme-event analysis. Station-based temporal records, such as WEATHER-5K~\cite{han2024weather}, complement gridded products by offering localized, long-term observational sequences for time-series forecasting and model evaluation.

In the hydrosphere, oceanographic time series are equally important. HYCOM~\cite{chassignet2007hycom} is more accurately described as a data-assimilative ocean modeling and reanalysis system rather than a pure observational dataset. It provides temporally evolving fields of ocean temperature, salinity, currents, and sea surface height, and is widely used for ocean circulation analysis, marine forecasting, and climate studies. Coastal records such as NOAA Tides and Currents~\cite{guest2006solutions} provide sea-level and tidal time series that are useful for storm-surge analysis, coastal hazard assessment, and erosion studies.

Time-series data are also central in the lithosphere. Global seismic waveform archives such as those provided by the IRIS Seismological Data Center~\cite{ahern2015iris}, along with benchmark datasets such as STEAD~\cite{mousavi2019stanford}, DiTing~\cite{zhao2023diting}, and Instance~\cite{michelini2021instance}, record ground motion at high temporal resolution for earthquake detection, phase picking, source characterization, and early warning. In addition, volcanic monitoring time series, including gas emissions, deformation signals, and thermal activity, support eruption forecasting and hazard assessment.

Overall, time-series data in Earth science are highly diverse in temporal scale, spatial support, and generation mechanism. Their importance lies not only in forecasting, but also in revealing process dynamics, supporting anomaly detection, and enabling physically grounded reasoning across Earth-system components.

\subsection{Evolution of Earth Science Models}

Earth science models have evolved substantially in scope, scale, and autonomy. A useful way to characterize this evolution is through three stages: \textbf{Scientific Perception}, \textbf{Scientific Reasoning}, and \textbf{Scientific Discovery}. As illustrated in Fig.~\ref{fig:taxonomy}, these stages are associated with a shift from task-specific models to general-purpose foundation models and, more recently, to agentic systems. Representative models across Earth-system spheres are summarized in Table~\ref{tab:sphere_models_all}.

In the \textbf{Scientific Perception} stage, models are developed primarily for predefined tasks in relatively narrow domains, such as classification, segmentation, retrieval, and spatiotemporal prediction. They are typically trained on specific modalities and optimized for fixed objectives, which yields strong task performance but limited transferability. The \textbf{Scientific Reasoning} stage is marked by the rise of large foundation models, especially LLMs and multimodal large models, which benefit from large-scale pretraining, instruction tuning, and improved multimodal alignment. These models support broader generalization, zero-shot transfer, and more flexible interaction with heterogeneous data and tools. The \textbf{Scientific Discovery} stage extends this trend toward autonomous or semi-autonomous systems that can plan workflows, invoke external tools, interact with simulations or databases, and iteratively refine outputs in support of scientific investigation.

Below, we review this evolution from a model--modality perspective, covering language, vision, multimodal models, and agentic systems.

\textbf{Language Foundation Models.}
Language-based modeling in Earth science has progressed from task-specific pre-trained language models to more general domain-oriented LLMs. As shown in Table~\ref{tab:sphere_models_all}, such models have already been applied across the atmosphere, hydrosphere, lithosphere, and cross-sphere settings.

\emph{Scientific Perception.}
Early pre-trained language models (PLMs) largely followed the pre-training-and-fine-tuning paradigm~\cite{devlin2019bert,yang2019xlnet,radford2018improving,radford2019language}. Representative examples include ClimateBERT~\cite{webersinke2021climatebert}, SpaBERT~\cite{li2022spabert}, and MGeo~\cite{ding2023mgeo}, which were designed for relatively narrow tasks such as climate-text classification, geographic entity linking, and query--POI matching. Their strengths lie in domain adaptation and efficient fine-tuning, but their applicability is constrained by task-specific objectives and limited training scope.

\emph{Scientific Reasoning.}
The emergence of large language models has substantially broadened the role of language modeling in Earth science. Models such as K2~\cite{deng2024k2}, OceanGPT~\cite{bi2023oceangpt}, GeoGalactica~\cite{lin2023geogalactica}, and ClimateChat~\cite{chen2025climatechat} are trained or adapted using broader corpora that may include geoscience papers, climate reports, encyclopedic resources, and scientific knowledge bases such as GAKG~\cite{deng2021gakg}. Compared with earlier PLMs, they support a wider range of tasks, including question answering, summarization, scientific dialogue, knowledge probing, and report generation. This broader language interface is particularly useful in Earth science, where knowledge is distributed across subfields with distinct terminologies, data conventions, and reasoning practices.

\textbf{Vision Foundation Models.}
Vision models in Earth science have developed from task-specific perceptual systems toward larger and more transferable visual foundation models. Here, ``vision'' includes a wide range of scientific visual inputs, including satellite and aerial imagery, radar products, underwater imagery, seismic sections, ecological imagery, and cryospheric visual fields.

\begin{table*}[ht]
\centering
\caption{Taxonomy of Earth science tasks. The classification is structured according to the specific Earth sphere and its corresponding data modalities and operational task types.}
\label{tab:tasks}
\resizebox{\textwidth}{!}{%
\begin{tabular}{|l|l|l|p{8cm}|}
\hline
\textbf{Sphere} & \textbf{Task Type} & \textbf{Data Type} & \textbf{Task Description} \\ \hline

\multirow{6}{*}{Atmosphere} 
& Weather Forecasting & Multimodal Met-Climate Series & Predicting medium- to long-term atmospheric states. \\ \cline{2-4} 
& Precipitation Nowcasting & High-resolution Radar/Sat Data & Ultra-short-term precipitation and extreme weather prediction. \\ \cline{2-4} 
& Climate Downscaling & Multimodal Climate Data & Enhancing spatial/temporal resolution of climate models. \\ \cline{2-4} 
& Model Bias Correction & Weather/Climate Series Data & Mitigating systematic errors in numerical models. \\ \cline{2-4} 
& Climate Textual Analysis & Domain-specific Text/Reports & Extracting insights from domain-specific climate texts. \\ \cline{2-4} 
& Atmospheric Pattern Recognition & Multimodal Weather Data & Identifying complex weather phenomena from grids. \\ \hline

\multirow{4}{*}{Lithosphere} 
& Geological Mapping \& Interpretation & Multispectral/Hyperspectral, DEMs & Delineating lithological units and structural features. \\ \cline{2-4} 
& Petrophysical \& Mineral Analysis & Spectral, Geophysical Data & Inferring rock properties and mineral compositions. \\ \cline{2-4} 
& Mineral Exploration \& Geohazards & Multi-source Geospatial Data & Predicting mineral prospectivity and assessing geohazards. \\ \cline{2-4} 
& Geo-environmental Assessment & RS, In-situ Observations & Evaluating surface and subsurface geological conditions. \\ \hline

\multirow{6}{*}{Hydrosphere} 
& Water Quality \& Dynamics Analysis & Multi-Spectral, Multi-Temporal & Monitoring water parameters and temporal dynamics. \\ \cline{2-4} 
& Streamflow Regionalization & Hydrometeorological Data & Transferring hydrological data to ungauged catchments. \\ \cline{2-4} 
& Hydrological Knowledge QA & Domain-specific Textual Data & Answering domain-specific hydrological questions via LLMs. \\ \cline{2-4} 
& Hydrological Feature Extraction & High-res Satellite Imagery & Segmenting water bodies like rivers and lakes. \\ \cline{2-4} 
& Extreme Hydro-event Forecasting & Time Series, Radar Data & Predicting floods, droughts, and extreme events. \\ \cline{2-4} 
& Marine Phenomenon Detection & Oceanographic, SAR Data & Tracking oceanic features like eddies and spills. \\ \hline

\multirow{7}{*}{Biosphere} 
& Species Distribution Modeling (SDM) & Occurrence Data, Env. Covariates & Predicting species geographic ranges. \\ \cline{2-4} 
& Range Mapping \& Dynamics & Spatial Distribution Data & Delineating and updating taxa boundaries. \\ \cline{2-4} 
& Global Carbon Budget Estimation & Biomass, Flux, and RS Data & Quantifying global carbon sinks and sources. \\ \cline{2-4} 
& Biodiversity Threat Assessment & Threat Data, LULC & Evaluating stressors on ecosystems and habitats. \\ \cline{2-4} 
& Phenotyping \& Trait Extraction & Imagery, Observational Data & Measuring morphological traits from biological data. \\ \cline{2-4} 
& Environmental Pollution Monitoring & Hyperspectral, In-situ Sensors & Tracking air, water, and soil pollutants. \\ \cline{2-4} 
& Crop Yield \& Growth Monitoring & Multispectral Time Series & Predicting agricultural yields and phenological stages. \\ \hline

\multirow{2}{*}{Cryosphere} 
& Sea Ice Dynamics Forecasting & SAR, Passive Microwave Data & Predicting sea ice extent and thickness. \\ \cline{2-4} 
& Glacier Mapping \& Mass Balance & Optical, SAR, Altimetry & Monitoring glacier boundaries and volume changes. \\ \hline

\multirow{3}{*}{Anthroposphere} 
& Urban Footprint \& Sprawl Monitoring & High-res Spatial Data & Mapping urban expansion and built-up areas. \\ \cline{2-4} 
& LULC Classification & Optical/SAR Multispectral & Categorizing land use and land cover classes. \\ \cline{2-4} 
& Disaster Damage Assessment & Pre/Post-event Imagery, Text & Evaluating post-disaster infrastructure and human impact. \\ \hline

\end{tabular}%
}
\end{table*}

\emph{Scientific Perception.}
Early vision models were largely built for specific tasks and modalities. Many focused on RGB or multispectral satellite imagery for scene classification, object detection, semantic segmentation, and change detection, supported by Earth observation data from sensors such as Landsat, Sentinel, and MODIS~\cite{masek2020landsat,phiri2020sentinel,justice2002overview}. Representative task-specific approaches include PhIRE~\cite{stengel2020adversarial}, ESRGAN-based downscaling~\cite{watson2020investigating}, TSTNet~\cite{hao2022two}, and TRS~\cite{zhang2021trs}.

At the same time, visual perception in Earth science has always involved more diverse sensing modalities. In meteorology, radar-based datasets such as SEVIR~\cite{veillette2020sevir}, MeteoNet~\cite{larvor2021meteonet}, and RainNet~\cite{chen2022rainnet} support precipitation estimation and nowcasting. In hydrospheric settings, visual models have been developed for flood understanding, water-body extraction, and marine scene analysis, as in FloodNet~\cite{rahnemoonfar2021floodnet}, MF-SegFormer~\cite{zhang2023water}, MarineDet~\cite{haixin2023marinedet}, and CoralSCOP~\cite{zheng2024coralscop}. In the cryosphere, visual spatiotemporal fields are used for sea-ice prediction and generation~\cite{xu2024sifm,jiang2024sicformer,finn2024towards,xu2024icediff}. In the lithosphere, seismic sections and subsurface imaging support interpretation and inversion, as in InversionNet~\cite{wu2019inversionnet} and DispFormer~\cite{liu2025dispformer}. Biosphere applications include vegetation and canopy analysis~\cite{pettorelli2013normalized,rege2024depth}. Across these settings, most models remained task-specific and were closely coupled to the sensing characteristics and physical structure of the target data.

\emph{Scientific Reasoning.}
Recent visual foundation models aim to learn more generalizable representations from large-scale Earth science data. In atmospheric and climate applications, representative models include ClimaX~\cite{nguyen2023climax}, WeatherGFM~\cite{zhao2024weathergfm}, Aurora~\cite{bodnar2024aurora}, and Prithvi WxC~\cite{schmude2024prithvi}. In broader Earth observation settings, models such as MMEarth~\cite{nedungadi2024mmearth} and EarthPT~\cite{smith2023earthpt} move toward unified pretraining across heterogeneous geospatial inputs.

At the same time, sphere-specific high-capacity models remain central in domains such as atmospheric forecasting, including FourCastNet~\cite{pathak2022fourcastnet}, GraphCast~\cite{lam2023graphcast}, PanGu-Weather~\cite{bi2022pangu}, and FengWu~\cite{chen2023fengwu}. More generally, visual foundation models are increasingly conditioned on metadata such as geolocation, acquisition time, sensor geometry, or auxiliary physical variables, reflecting the structured nature of Earth science observations. Generative modeling is also becoming more prominent, with diffusion-based approaches being used for super-resolution, interpolation, reconstruction, and probabilistic forecasting under scientific constraints.

\textbf{Multimodal Foundation Models.}
A major recent trend is the move from unimodal systems toward models that jointly process imagery, text, metadata, geolocation, and temporal context. This shift is particularly consequential in Earth science, where interpretation often depends on combining heterogeneous evidence sources. As shown in Fig.~\ref{fig:taxonomy} and Table~\ref{tab:sphere_models_all}, multimodal models now occupy a central position in the reasoning stage.

\emph{CLIP-style models.}
One line of work follows the CLIP paradigm~\cite{radford2021learning}, using dual encoders to align images and text through contrastive learning. In remote sensing and Earth observation, this approach has supported zero-shot classification, retrieval, and representation learning, as in SkyScript~\cite{wang2024skyscript}, RS5M~\cite{zhang2023rs5m}, and S-CLIP~\cite{mo2023s}. Some variants further incorporate geographic priors or structured metadata.

\emph{Generative multimodal models.}
A second line of work uses conditional diffusion and related generative frameworks to synthesize imagery from text, metadata, or partial observations~\cite{sebaq2024rsdiff,khanna2312diffusionsat}. In Earth science, these models are relevant not only for generation, but also for reconstruction, forecasting, and data completion in settings with sparse or irregular observations.

\emph{Multimodal large language models.}
A third direction combines visual encoders with LLMs to build MLLMs for Earth science understanding and reasoning. Representative examples include LHRS-Bot~\cite{muhtar2024lhrs}, EarthGPT~\cite{zhang2024earthgpt}, TEOChat~\cite{irvin2024teochat}, GeoChat~\cite{kuckreja2024geochat}, and CLLMate~\cite{li2024cllmate}. These systems support multimodal question answering, visual grounding, report generation, and event understanding. As reflected in Table~\ref{tab:sphere_models_all}, multimodal modeling is expanding from atmospheric remote sensing to marine science, lithosphere-oriented applications, and cross-sphere Earth observation.

\textbf{Agentic Systems and Scientific Discovery.}
The most recent stage in this evolution is the emergence of agentic systems, in which foundation models serve not only as predictors or reasoners, but also as coordinators of broader scientific workflows. Compared with standard multimodal models, agents can decompose tasks, select tools, query databases, invoke simulation modules, and integrate intermediate outputs to support more complex objectives.

Although still at an early stage, initial examples already suggest the potential of this direction, such as PANGAEA GPT~\cite{pantiukhin2025accelerating}, Earth-Agent~\cite{feng2025earth}, GeoFactory~\cite{chen2025geofactory}, and \emph{From News to Forecast}~\cite{wang2024news} illustrate emerging capabilities in autonomous scientific assistance, multimodal workflow orchestration, and open-ended geoscientific task coordination.
These systems are particularly relevant to Earth science because many research workflows require iterative interaction among data archives, physical simulators, domain constraints, and expert interpretation. Accordingly, effective agentic systems must support not only reasoning, but also reliable tool use, uncertainty management, and scientific traceability.

%% file: draft/033_sphere_models.tex
\section{Domain-Specific Earth FMs}
\label{sec:domain_fms}

The architecture of AI-driven Earth system research is fundamentally built upon the mastery of individual environmental domains. Before achieving holistic, cross-sphere unification, foundation models must first decode the highly specialized physical laws, complex spatiotemporal dynamics, and distinct multimodal data distributions inherent to isolated Earth spheres. Domain-specific Earth Foundation Models (FMs) are explicitly engineered to operate within these boundaries. They represent a critical evolutionary step beyond computationally expensive numerical physics, leveraging massive domain-specific datasets—such as high-resolution atmospheric reanalysis grids or anthropospheric trajectory logs—to deliver unprecedented predictive accuracy, computational efficiency, and localized analytical depth. This section systematically reviews these specialized models by partitioning them according to their primary Earth spheres. For each domain, we explore the fundamental data modalities, categorize the core predictive and analytical tasks, and evaluate the real-world application paradigms that these focused models enable, thereby laying the critical groundwork for the unified systems discussed later in this survey.

 \begin{figure*}[h]
	\centering
	\includegraphics[width=1.0\textwidth]{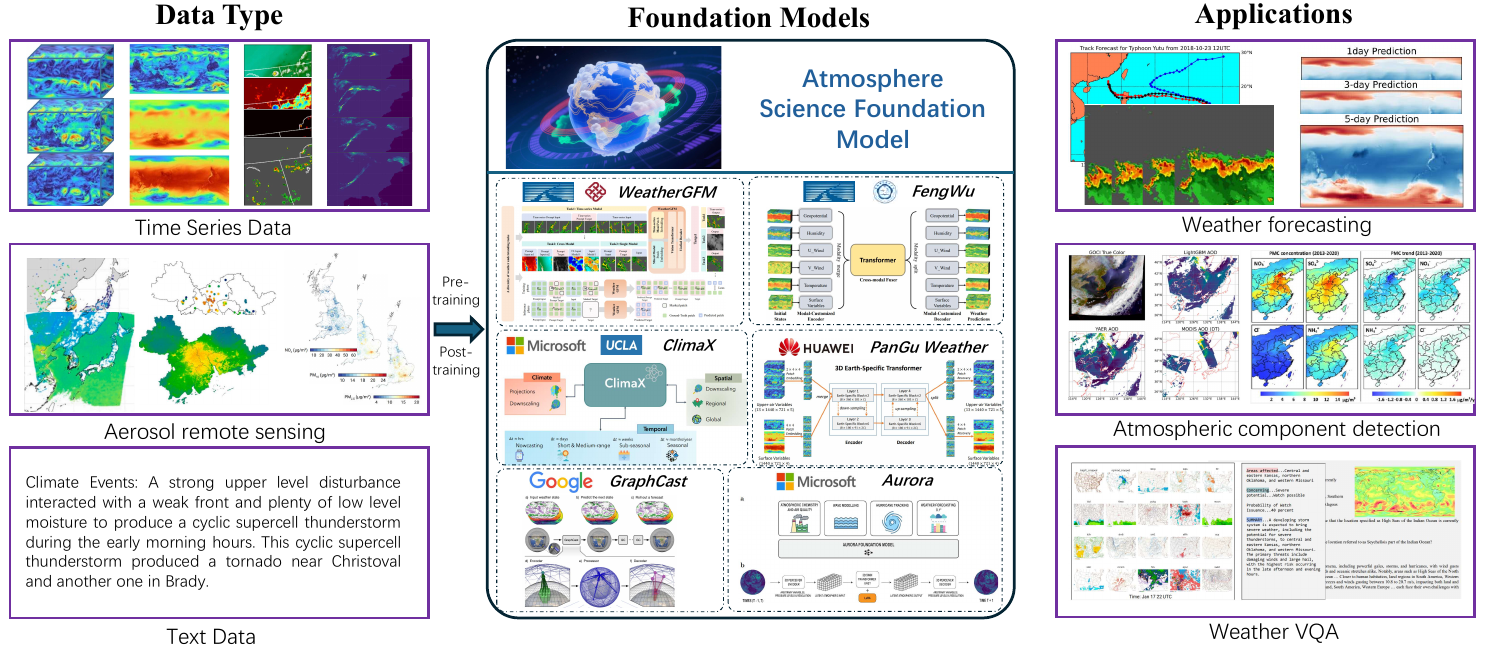}
        \caption{Illustration of atmosphere science data type, foundation models and applications.
        }
        \label{atmos_case}
\end{figure*}


\subsection{Atmosphere Science}
\label{sec:mllm_train}

At the application level, we divide the existing literature into three main categories based on specific data categories: Vision foundation models, LLM-powered agent and MLLM-powered agent.

\subsubsection{Data}
Investigations into the atmosphere typically necessitate the exploration of diverse spatiotemporal and textual data, which can be broadly categorized into meteorological variables, climate text data, and multimodal resources. Within meteorological variables, station-based observation data originate from global weather stations, collecting high-resolution measurements (e.g., temperature, humidity, wind speed, and precipitation) at specific locations. Datasets like WEATHER-5K~\cite{han2024weather} and Digital Typhoon~\cite{kitamoto2023digital} exemplify this, supporting the extraction of detailed insights into local weather patterns and extreme events. However, because station coverage is often uneven, with sparse distribution in remote regions such as oceans, mountains, and deserts, gridded reanalysis data offers a complementary global view. Derived from a combination of station observations, satellite measurements, and numerical weather prediction (NWP) models, gridded datasets such as ERA5~\cite{rasp2020weatherbench}, CMIP6~\cite{rasp2024weatherbench}, and CRA5~\cite{han2024cra5} provide consistent spatial coverage for large-scale climate modeling. Beyond physical variables, textual data has become increasingly vital for atmospheric AI. Datasets containing textual claim pairs, such as CLIMATE-FEVER~\cite{diggelmann2020climate}, and topic annotations like ClimaText~\cite{varini2020climatext}, enable large language models to perform specialized tasks including fact-checking, named entity extraction, and public stance detection on climate policies. Finally, multimodal atmospheric data bridges the gap between unstructured text and physical grids. Resources like WeatherQA~\cite{ma2024weatherqa} and SEVIR~\cite{veillette2020sevir} integrate reanalysis grids or meteorological imagery with textual question-answering pairs, providing the critical foundation for training MLLMs to perform complex meteorological reasoning, visual forecasting, and decision-making.

\subsubsection{Tasks}

For atmospheric science, we dissect the existing literature into four primary application domains: Spatiotemporal Forecasting, Textual Reasoning \& QA, Multimodal Understanding, and Agentic Workflows. The forecasting category further encompasses specific physical subtasks such as Climate Forecasting, Subseasonal to Medium-Range Forecasting, Nowcasting, Downscaling, Data Assimilation, and Bias Correction. Conversely, textual and multimodal capabilities are treated as generalized subjects (e.g., Climate Text Analysis), as their specific task characteristics, such as relationship extraction, visual question answering, or stance detection, are typically delineated by downstream datasets rather than intrinsic architectural differences in the foundation models themselves.

\subsubsection{Applications}

The integration of Earth FMs into atmospheric applications is driving a paradigm shift from traditional numerical weather prediction (NWP) to data-driven, AI-assisted meteorology. These models capture complex, non-linear interactions within atmospheric systems, providing breakthroughs in prediction speed, spatial resolution, and semantic reasoning.

\textbf{Spatiotemporal Forecasting \& Simulation:} This domain represents the most mature application of Earth FMs, evolving from single-task deterministic predictions to unified, probabilistic, and physics-informed simulations. Early foundation models exploited Vision Transformers (ViT) and Graph Neural Networks (GNN); for instance, FourCastNet~\cite{pathak2022fourcastnet} and GraphCast~\cite{lam2023graphcast} delivered impressive global forecasts at a $0.25^\circ$ resolution. For ultra-large-scale medium-term forecasting, PanGu-Weather~\cite{bi2022pangu} utilizes 3D Earth-specific transformers for rapid predictions across 13 vertical levels, while FengWu~\cite{chen2023fengwu} incorporates cross-modal fusion transformers supervised by an uncertainty loss mechanism. FuXi~\cite{chen2023fuxi} cascades cubic embeddings to achieve 15-day forecast performance comparable to the ECMWF ensemble model.

Recent advancements focus on unified, all-purpose architectures. ClimaX~\cite{nguyen2023climax} introduces variable tokenization to fuse diverse weather variations, enabling adaptation across global forecasting and local downscaling. Similarly, Prithvi WxC~\cite{schmude2024prithvi} and WeatherGFM~\cite{zhao2024weathergfm} address multi-task limitations, with the latter unifying up to 10 distinct meteorological tasks via visual prompting. Aurora~\cite{bodnar2024aurora}, a 1.3B-parameter model, outperforms operational forecasts on specialized tasks like atmospheric chemistry and extreme event tracking.

To bridge the gap between data-driven black boxes and physical conservation laws, hybrid and probabilistic frameworks have emerged. NeuralGCM~\cite{kochkov2024neural} seamlessly couples machine learning-based sub-grid parameterizations with traditional fluid dynamics solvers. GenCast~\cite{price2025gencast} introduced diffusion-based generative ensembles, a trajectory further advanced by architectures like WeatherNext 2, which injects noise directly into the model's function space to guarantee physically realistic outputs for highly localized, extreme events. Furthermore, to address global data sparsity, frameworks such as ESFM-MoE~\cite{cheng2026esfm} utilize climate-semantic routing to harmonize heterogeneous observations across varying resolutions. Finally, self-supervised approaches like W-MAE~\cite{man2023w} continue to leverage masked autoencoders to extract highly robust spatiotemporal representations without explicit task supervision.

\textbf{Textual Reasoning \& QA:} Beyond numeric grids, atmospheric research requires the semantic analysis of climate policies, scientific literature, and public discourse. Domain-specific language models such as ClimateBert~\cite{webersinke2021climatebert} and CliMedBert~\cite{fard2022climedbert} pioneered climate text analysis, focusing on fact-checking, risk disclosure, and entity extraction. Building upon this, ClimateGPT~\cite{thulke2024climategpt} provides an advanced Retrieval-Augmented Generation (RAG) framework tailored for complex, interdisciplinary climate discourse. For highly specialized weather reasoning, ClimateLLM~\cite{li2025climatellm} proposes a physics-aligned frequency-domain perception framework. By combining Fourier-based frequency decomposition with dynamic LLM prompting, it effectively separates energy amplitude from spatial propagation, achieving vastly superior performance in modeling and answering complex queries regarding volatile extreme weather events.

\textbf{Multimodal Understanding:} To close the cognitive gap between computing meteorological variables and generating actionable textual narratives, Multimodal Large Language Models (MLLMs) are being developed to process numerical data, satellite imagery, and observational reports simultaneously. CLLMate~\cite{li2025cllmate} introduces a multimodal framework aligning raster meteorological data (like ERA5) with textual narratives for Weather and Climate Event Forecasting (WCEF). Similarly, RadarQA~\cite{he2025radarqa} pairs visual environmental factors, such as radar reflectivity, with expert forecast analyses to rigorously benchmark visual reasoning capabilities. Advancing architectural integration, Aquilon~\cite{varambally2025aquilon} allows MLLMs to reason over complex weather data by translating dense embeddings directly from predictive foundation models into LLM-compatible visual tokens. Unifying these parallel tracks, Omni-Weather~\cite{zhou2025omni} expresses both generative forecasting (e.g., radar nowcasting) and diagnostic reasoning in a standardized sequence-to-sequence format, demonstrating that training on both tasks provides mutually reinforcing supervision signals that sharpen both predictive and semantic outputs.

\textbf{Agentic Workflows:} The most advanced frontier of atmospheric AI lies in autonomous agents capable of dynamic interaction, self-reflection, and tool utilization. Some frameworks enhance forecasting by integrating unstructured societal data; the ``From News to Forecast'' approach~\cite{wang2024news} deploys LLM-based agents to continuously filter global news databases and align real-world societal events with historical time-series fluctuations, enabling highly context-aware predictions. Pushing further into scientific workflow automation, the ZEPHYRUS framework and its associated agentic tools (like Anemoi)~\cite{varamballyanemoi} act as an autonomous AI co-scientist. Equipped with deterministic tools to securely query 4D NetCDF files, run regional statistical simulations, and refine its hypotheses through iterative conversational loops, these agents effectively democratize access to complex, supercomputer-grade meteorological analysis.

\begin{table*}[t]
  \centering
  \caption{Summary of representative Earth science datasets and their applications across different Earth spheres.}
  \label{tab:dataset_all}
  \resizebox{\textwidth}{!}{
    \begin{tabular}{lllll}
      \toprule
      Sphere        & Data Type                                   & Dataset                                                      & Timeframe       & Applications                                             \\
      \midrule

      \multicolumn{5}{l}{\textbf{A. Atmosphere}} \\
      & Gridded climate model data           & CMIP6~\cite{rasp2020weatherbench,rasp2024weatherbench}                 & 1850–2100     & Forecasting; Projection; Downscaling; Bias correction; Data assimilation \\
      & Gridded reanalysis data              & ERA5~\cite{rasp2020weatherbench,rasp2024weatherbench} / CRA5~\cite{han2024cra5} & 1779–present  & Forecasting; Projection; Downscaling; Bias correction; Data assimilation \\
      & Gridded reanalysis data              & ExtremeWeather~\cite{racah2017extremeweather}        & 1979–2005     & Forecasting; Projection; Nowcasting                      \\
      & Gridded model \& reanalysis          & ClimateNet~\cite{prabhat2020climatenet}            & 1996–2010     & Forecasting; Projection; Weather‐pattern understanding  \\
      & Gridded ensemble forecast            & ENS‐10~\cite{ashkboos2022ens}                & 1998–2017     & Forecasting; Downscaling; Bias correction; Data assimilation \\
      & Gridded model data                   & ClimART~\cite{cachay2021climart}               & 1979–2014     & Forecasting; Projection; Bias correction; Data assimilation; Weather‐pattern understanding \\
      & Meteorological time-series           & WEATHER-5K~\cite{han2024weather}               & 2014-2023     & Forecasting \\
      & Meteorological observations          & Digital Typhoon~\cite{kitamoto2023digital}       & 1978–2022     & Forecasting; Pattern understanding; Nowcasting         \\
      & Radar \& gauge observations          & IowaRain~\cite{sit2021iowarain}              & 2016–2019     & Forecasting; Projection; Pattern understanding         \\
      & Radar observations                   & \href{https://tianchi.aliyun.com/competition/entrance/231662/information}{SRAD2018}              & 2018          & Pattern understanding; Nowcasting                      \\
      & Air quality observations             & KnowAir~\cite{wang2020pm2}               & 2015–2018     & Forecasting                                            \\
      & Satellite observations               & NASA~\cite{chen2023prompt}       & 2012–2022     & Forecasting                                            \\
      & Gridded climate observations         & PRISM~\cite{nguyen2023climatelearn}                & 1895–present  & Forecasting; Projection; Downscaling; Bias correction; Data assimilation; Pattern understanding \\
      & Radar observations                   & RainNet~\cite{chen2022rainnet}               & 2002—2018      & Downscaling; Pattern understanding; Nowcasting         \\
      & Multimodal obs. \& reanalysis          & EarthNet2021~\cite{requena2021earthnet2021}         & 2018          & Forecasting; Projection; Pattern understanding         \\
      & Multimodal meteorological data       & KoMet~\cite{kim2022benchmark}                & 2011–2018     & Forecasting                                            \\
      & Multimodal meteorological data       & PostRainBench~\cite{tang2023postrainbench}  & 2020–2021     & Forecasting; Pattern understanding; Nowcasting         \\
      & Multimodal meteorological data       & MeteoNet~\cite{larvor2021meteonet}              & 2016–2018     & Forecasting; Pattern understanding; Nowcasting         \\
      & Multimodal meteorological data       & RAIN‐F+~\cite{choi2021rain}              & 2017–2019     & Forecasting; Pattern understanding; Nowcasting         \\
      & Multimodal obs. \& reanalysis          & RainBench~\cite{de2021rainbench}             & 2000–2017     & Forecasting; Projection; Downscaling; Pattern understanding; Nowcasting \\
      & Multimodal meteorological data       & Weather2K~\cite{zhu2023weather2k}             & 2017–2021     & Forecasting                                            \\
      & Multimodal obs. \& reanalysis          & LSDSSIMR~\cite{bai2023lsdssimr}              & 2020–2022     & Forecasting; Projection; Pattern understanding         \\
      & Textual claim \& stance pairs          & CLIMATE-FEVER~\cite{diggelmann2020climate} / Climate-Stance~\cite{vaid2022towards} & — & Fact-checking; Claim verification; Public stance detection \\
      & Textual IE \& entity annotations       & ClimateBERT-NetZero~\cite{schimanski2023climatebert} / SCIDCC~\cite{mishra2021neuralnere} & — & Named entity/relation extraction; ESG analysis; KG construction \\
      & Textual topic annotations              & ClimaText~\cite{varini2020climatext} & — & Climate topic detection; Sentence-level classification \\
      & Textual QA \& benchmark suites         & ClimateEval~\cite{kurfali2025climateeval} / ClimaQA~\cite{manivannan2024climaqa} & — & Question answering; Comprehensive LLM capability evaluation \\
      & Reanalysis \& textual QA pairs         & WeatherQA~\cite{ma2024weatherqa} / ClimateIQA~\cite{chen2025climateiqa} / CLLMate~\cite{li2025cllmate} & — & Climate Text Analysis \\
      & Multimodal meteorological imagery    & SEVIR~\cite{veillette2020sevir}                & —             & Forecasting; Projection; Data assimilation; Pattern understanding; Nowcasting \\
      & Reanalysis \& textual QA pairs         & RadarQA~\cite{he2025radarqa}            & —             & Weather Forecast Analysis    \\
      & Reanalysis \& textual CoT pairs        & OMNI-WEATHER CoT~\cite{zhou2025omni}            & —             & Weather Generation and Understanding    \\

      \midrule
      \multicolumn{5}{l}{\textbf{B. Hydrosphere}} \\
      & Domain-specific textual data           & WaterER~\cite{xu2024unlocking}    & — & Water engineering and research \\
      & UAV imagery \& text pairs              & FloodNet~\cite{rahnemoonfar2021floodnet}                    & — & Post‐flood scene understanding (VQA)     \\
      & Underwater imagery \& QA pairs         & J-EDI QA (GODAC)~\cite{yoshida2024j}                                 & — & Deep‐sea organism question answering       \\
      & Multi-sensor EO imagery                & EarthMind-Bench~\cite{shu2025earthmind}             & — & Remote‐sensing perception→reasoning; multitask \\
      
      \midrule
      \multicolumn{5}{l}{\textbf{C. Biosphere}} \\
      & Species occurrence records             & GBIF Database~\cite{telenius2011biodiversity}        & - & Global biodiversity occurrence integration \\
      & Ecological interaction networks        & GloBI~\cite{noori2026curated} & - & Species‐interaction/eco‐network analysis \\
      & Multimodal EO \& text pairs            & REO-Instruct~\cite{xue2025towards} & - & Forest ecological analysis \\
      & Species distribution \& observation    & SRMs dataset~\cite{hamilton2024combining} / eBird Status \& Trends~\cite{eBirdStatusTrends2023} & - & Species range maps \\

      \midrule
      \multicolumn{5}{l}{\textbf{D. Lithosphere}} \\
      & Multispectral imagery                  & Landsat‐9~\cite{masek2020landsat}        & 1972–present & Lithology mapping; mineral exploration; time‐series analysis \\
      & Multispectral imagery                  & Sentinel-2~\cite{phiri2020sentinel}            & 2015–present & Lithology mapping; mineral exploration    \\
      & Multispectral imagery                  & ASTER~\cite{baldridge2009aster}                  & 2000–present & Hydrothermal alteration mineral mapping   \\
      & VHR multispectral imagery              & WorldView-2/GeoEye~\cite{aguilar2013geoeye}/Pleiades~\cite{gleyzes2012pleiades}/Gaofen-2~\cite{tong2018large} & various & Regional lithology mapping; fine‐feature ID \\
      & Airborne hyperspectral imagery         & AVIRIS~\cite{xiao2004using}, HyMap~\cite{schlerf2005remote}, HySpex~\cite{jiang2020exploring}             & various & Ultra‐high‐res. lithology/mineral mapping  \\
      & Spaceborne hyperspectral imagery       & Hyperion~\cite{datt2003preprocessing}/GF-5 AHSI~\cite{ye2020application}/EMIT~\cite{green2023performance}   & 2000–present & Surface mineral spectroscopy; compositional analysis \\
      & Seismic waveforms \& records           & STEAD~\cite{mousavi2019stanford} / DiTing~\cite{zhao2023diting} / Instance~\cite{michelini2021instance} & 1970-present & Earthquake monitoring; early warning; subsurface tomography; seismic hazard analysis \\
      & Gravimetric \& magnetic grids          & GRACE~\cite{tapley2004grace} / GOCE~\cite{brockmann2014egm_tim_rl05} / Swarm~\cite{friis2006swarm} / EMAG2~\cite{maus2009emag2} & 2002–present & Lithospheric structure; crustal thickness estimation; resource exploration; tectonic analysis \\
      & Seismic simulation \& tomographic data & OpenSWI~\cite{liu2025openswi} / OpenFWI~\cite{deng2022openfwi} / GlobalTomo~\cite{li2024globaltomo} / CigFacies~\cite{gao2024cigfacies} & various & Subsurface imaging; resource exploration; geological hazard assessment (e.g. active fault) \\
      & Geochemical \& petrological records    & EarthChem~\cite{walker2005earthchem} / GEOROC~\cite{digis2021georoc} / PetDB~\cite{lehnert2000global} & various & Resource exploration; earth evolution studies; environmental geochemistry \\

      \midrule
      \multicolumn{5}{l}{\textbf{E. Anthroposphere}} \\
      & RS image–text pairs               & RemoteCLIP~\cite{liu2024remoteclip} & - & Retrieval; object detection; segmentation \\
      & RS image–text pairs               & SkyCLIP~\cite{wang2024skyscript} & - & Classification; detection; segmentation \\
      & RS image–text pairs               & S-CLIP dataset~\cite{mo2023s} & - & Representation learning; classification \\
      & RS image–text pairs               & RS5M (GeoRSCLIP)~\cite{zhang2024rs5m} & - & Vision–language pretraining \\
      & RS image–text pairs               & RS-CLIP dataset~\cite{li2023rs} & - & Representation learning \\
      & Geo-tagged multispectral text pairs    & S2-100K~\cite{klemmer2025satclip} & 2010–present & Geo-localization; multi-task classification \\
      & RS visual instruction / dialogue pairs & SkyEye-968k~\cite{zhan2025skyeyegpt} & - & Instruction tuning; visual QA; dialogue \\
      & RS expert-annotated caption pairs      & RSICap~\cite{hu2025rsgpt} & - & Captioning; visual reasoning \\
      & RS multimodal QA pairs                 & MMRS-1M~\cite{zhang2024earthgpt} & - & Instruction tuning; multimodal QA \\
      & RS image–text instruction pairs        & LHRS-Align/Instruct~\cite{muhtar2024lhrs} & - & Instruction fine-tuning \\
      & RS visual QA pairs                     & RSIVQA~\cite{lobry2020rsvqa} & - & Visual QA \\

      \bottomrule
    \end{tabular}
  }
\end{table*}

\begin{table*}[htbp]
  \centering
  \caption{Summary of representative Earth science foundation models and their specific tasks across different Earth spheres, including unified Earth FMs. The table details the category, methodology, domain, base architecture, and institutional origin.}
  \resizebox{1\textwidth}{!}{ 
    \begin{tabular}{c|c|ccc|cc} 
    \toprule
    \textbf{Sphere}        & \textbf{Category (Modality \& Task)} & \textbf{Method}                                      & \textbf{Task/Domain}                               & \textbf{Base}            & \textbf{Institute}     & \textbf{Year} \\
    \midrule
    
    \multirow{21}{*}{Atmosphere}
      & \multirow{12}{*}{Spatiotemporal Forecasting}
        & FourCastNet~\cite{pathak2022fourcastnet}             & Forecasting                                        & RNN             & Nvidia        & 2022 \\
      &                                 & GraphCast~\cite{lam2023graphcast}                    & Forecasting                                        & GNN             & Google        & 2022 \\
      &                                 & PanGu‐Weather~\cite{bi2022pangu}                  & Forecasting                                        & Transformer     & HUAWEI        & 2023 \\
      &                                 & FengWu~\cite{chen2023fengwu}                         & Forecasting                                        & Transformer     & SHAI          & 2023 \\
      &                                 & FuXi~\cite{chen2023fuxi}                             & Forecasting                                        & Transformer     & FuDan         & 2023 \\
      &                                 & W‐MAE~\cite{man2023w}                                & Forecasting                                        & Transformer     & UESTC         & 2023 \\
      &                                 & Aurora~\cite{bodnar2024aurora}                       & Forecasting                                        & Transformer     & Microsoft     & 2024 \\
      &                                 & ClimaX~\cite{nguyen2023climax}                       & Forecasting/Downscaling                            & Transformer     & UCLA          & 2023 \\
      &                                 & Prithvi WxC~\cite{schmude2024prithvi}                & Forecasting/Downscaling                            & Transformer     & IBM           & 2024 \\
      &                                 & WeatherGFM~\cite{zhao2024weathergfm}                 & Forecasting/Downscaling/Nowcasting                 & Transformer     & SHAI          & 2024 \\
      &                                 & NeuralGCM~\cite{kochkov2024neural}                   & Hybrid Weather/Climate Simulation                  & ML+Physics      & Google        & 2024 \\
      &                                 & GenCast~\cite{price2025gencast}                      & Probabilistic Weather Forecasting                  & Transformer     & DeepMind      & 2024 \\
    \cmidrule{2-7} 
      & \multirow{4}{*}{Textual Reasoning \& QA}
        & ClimateBert~\cite{webersinke2021climatebert}         & Climate Text Analysis                              & Transformer     & FAU           & 2021 \\
      &                                 & CliMedBert~\cite{fard2022climedbert}                 & Climate Text Analysis                              & Transformer     & UNMC          & 2022 \\
      &                                 & ClimateGPT~\cite{thulke2024climategpt}               & Climate Text Analysis                              & Transformer     & AppTek        & 2025 \\
      &                                 & ClimateLLM~\cite{li2025climatellm}                   & Forecasting QA                                     & Transformer     & USC           & 2025 \\
    \cmidrule{2-7}
      & \multirow{3}{*}{Multimodal Understanding}
        & CLLMate~\cite{li2025cllmate}                         & Multimodal Understanding                           & Transformer     & HKST          & 2024 \\
      &                                 & RadarQA~\cite{he2025radarqa}                         & Multimodal Understanding                           & Transformer     & SHAI          & 2025 \\
      &                                 & Aquilon~\cite{varambally2025aquilon}                 & Multimodal Understanding                           & Qwen-VL         & HDSI          & 2025 \\
    \cmidrule{2-7}
      & \multirow{2}{*}{Agentic Workflows}
        & From News to Forecast~\cite{wang2024news}            & Automated Forecasting Agent                        & Agent           & USYD          & 2024 \\
      &                                 & ZEPHYRUS~\cite{varamballyanemoi}                     & Multimodal Weather Agent                           & Agentic         & UCSD          & 2025 \\
    \midrule
    
    \multirow{13}{*}{Hydrosphere}
      & \multirow{2}{*}{Spatiotemporal Forecasting}
        & KUNPENG~\cite{wang2024kunpeng}                       & Ocean Environmental Forecasting                    & Transformer     & SHOU          & 2024 \\
      &                                 & IBM Ocean FM~\cite{dawson2025sentinel}               & Ocean Dynamics \& Global Tracking                  & Transformer     & IBM           & 2024 \\
    \cmidrule{2-7}
      & \multirow{3}{*}{Textual Reasoning \& QA}
        & OceanGPT~\cite{bi2023oceangpt}                       & QA / Ocean Science                                 & Transformer     & ZJU           & 2023 \\
      &                                 & WaterGPT~\cite{ren2024watergpt}                      & QA / Hydrology                                     & Transformer     & XDU       & 2024 \\
      &                                 & Llamarine~\cite{nguyen2025llamarine}                 & Maritime Navigation Reasoning                      & Transformer     & Aitomatic  & 2025 \\
    \cmidrule{2-7}
      & \multirow{3}{*}{Multimodal Understanding}
        & MarineGPT~\cite{zheng2023marinegpt}                  & Vision–Language Understanding                      & Transformer     & HKUST         & 2023 \\
      &                                 & CoralVQA~\cite{han2025coralvqa}                      & Coral Visual QA                                    & Transformer     & BUPT          & 2024 \\
      &                                 & LITE~\cite{li2024lite}                               & Multimodal Ecosystem Model                         & Transformer     & HRLICS        & 2024 \\
    \cmidrule{2-7}
      & \multirow{5}{*}{Visual Perception}
        & CoralSCOP~\cite{zheng2024coralscop}                  & Coral Image Segmentation                           & Transformer     & HKST          & 2024 \\
      &                                 & MarineDet~\cite{haixin2023marinedet}                 & Marine Object Detection                            & Transformer     & HKST          & 2023 \\
      &                                 & MarineInst~\cite{zheng2024marineinst}                & Instance Segmentation                              & Transformer     & ECVA          & 2023 \\
      &                                 & MarineSaliency~\cite{chang2025marine}                & Saliency Segmentation                              & Diffusion       & WHU          & 2025 \\
      &                                 & MF-SegFormer~\cite{zhang2023water}                   & Water‐Body Extraction                              & Transformer     & XDU     & 2023 \\
    \midrule
    
    \multirow{4}{*}{Biosphere}
      & \multirow{2}{*}{Textual Reasoning \& QA}
        & Chatecology~\cite{yangchatecology}                   & Green‐Tide QA                                      & Transformer     & NMEFC          & 2024 \\
      &                                 & BioRAG~\cite{wang2024biorag}                         & Molecular Biology QA                               & Transformer     & CAS    & 2024 \\
    \cmidrule{2-7}
      & \multirow{1}{*}{Multimodal Understanding}
        & Le‐SINR~\cite{hamilton2024combining}                 & Species Range Estimation                           & Transformer     & UMass         & 2024 \\
    \cmidrule{2-7}
      & \multirow{1}{*}{Visual Perception}
        & DepthAnyCanopy~\cite{rege2024depth}                  & Canopy Height Estimation                           & Transformer     & POLITO         & 2024 \\
    \midrule
    
    \multirow{4}{*}{Cryosphere}
      & \multirow{2}{*}{Spatiotemporal Forecasting}
        & SIFM~\cite{xu2024sifm}                               & Subseasonal Ice Prediction                         & CNN+RNN         & SHAI          & 2024 \\
      &                                 & SICFormer~\cite{jiang2024sicformer}                  & Ice Concentration Prediction                       & 3D-Swin         & CAS           & 2024 \\
    \cmidrule{2-7}
      & \multirow{2}{*}{Multimodal Generation}
        & SeaIceDiffusion~\cite{finn2024towards}               & Sea-Ice Discovery                                  & Diffusion       & CEREA     & 2024 \\
      &                                 & IceDiff~\cite{xu2024icediff}                         & High-Res Sea-Ice Forecasting                       & Diffusion       & FuDan       & 2024 \\
    \midrule
    
    \multirow{21}{*}{Anthroposphere}
      & \multirow{6}{*}{Spatiotemporal Forecasting}
        & ST-LLM~\cite{liu2024st}                              & Traffic Forecasting (Spatio-temporal)              & Transformer     & PKU & 2024 \\
      &                                 & TPLLM~\cite{ren2024tpllm}                            & Traffic Forecasting (Spatio-temporal)              & Transformer     & BUAA       & 2024 \\
      &                                 & UrbanGPT~\cite{li2024urbangpt}                       & Data-scarce Forecasting                            & Transformer     & HKU & 2024 \\
      &                                 & TrafficSafetyGPT~\cite{zheng2023trafficsafetygpt}    & Traffic Accident Prediction                        & Transformer     & UCF           & 2023 \\
      &                                 & DeepAir~\cite{zhao2023deepair}                       & Air Quality Forecasting                            & CNN+Transformer & BU         & 2023 \\
      &                                 & AirFormer~\cite{liang2023airformer}                  & Air Quality Forecasting                            & Transformer     & NUS & 2023 \\
    \cmidrule{2-7}
      & \multirow{6}{*}{Textual Reasoning \& QA}
        & UrbanPlanBench~\cite{zheng2025urbanplanbench}        & Planning Knowledge QA                              & Transformer     & THU  & 2025 \\
      &                                 & TransitGPT~\cite{devunuri2025transitgpt}             & NL Query Interface for Transit                     & Transformer     & UIUC           & 2025 \\
      &                                 & TrafficGPT~\cite{zhang2024trafficgpt}                & NL Query Interface for Traffic                     & Transformer     & XJTU & 2024 \\
      &                                 & CityGPT~\cite{guan2024citygpt}                       & NL Query Interface for Cities                      & Transformer     & THU      & 2024 \\
      &                                 & ChatLaw~\cite{cui2023chatlaw}                        & Legal Assistance                                   & Transformer     & PKU & 2023 \\
      &                                 & WatchoverGPT~\cite{shahid2024watchovergpt}           & Emergency Response                                 & Transformer     & SCUT          & 2024 \\
    \cmidrule{2-7}
      & \multirow{1}{*}{Multimodal Understanding}
        & PlanGPT-VL~\cite{zhu2025plangpt}                     & Doc Generation \& Evaluation                       & Transformer     & PKU & 2025 \\
    \cmidrule{2-7}
      & \multirow{6}{*}{Agentic Workflows \& Control}
        & City-LEO~\cite{jiao2024city}                         & Decision Making under Uncertainty                  & Transformer     & BTBU  & 2024 \\
      &                                 & Open-TI~\cite{da2024open}                            & Demand Optimization \& Signal Ctrl.                & Transformer     & SCAI           & 2024 \\
      &                                 & LA-Light~\cite{wang2024llm}                          & Traffic Signal Control (hybrid)                    & Trans.+Tool     & CUHK & 2024 \\
      &                                 & iLLM-TSC~\cite{pang2407illm}                         & Signal Control (RL-assisted)                       & Trans.+RL       & CUHK           & 2024 \\
      &                                 & CoLLMLight~\cite{yuan2025collmlight}                 & Multi-agent Signal Optimization                    & Trans.+GNN      & HKST           & 2025 \\
      &                                 & LLM4DistReconfig~\cite{christou2025llm4distreconfig} & Power‐Grid Reconfiguration Agent                   & Transformer     & NYU & 2025 \\
    \midrule
    
    \multirow{11}{*}{Lithosphere}
      & \multirow{2}{*}{Textual Reasoning \& QA}
        & SeisMoLLM~\cite{wang2025seismollm}                   & Earthquake Monitoring QA                           & Transformer     & SHAI         & 2025 \\
      &                                 & SeisGPT~\cite{meng2024seisgpt}                       & Building Damage Assessment                         & Transformer     & Tongji  & 2024 \\
    \cmidrule{2-7}
      & \multirow{2}{*}{Multimodal Understanding}
        & MineAgent~\cite{yu2024mineagent}                     & RS Mineral Exploration                             & MLLM            & UTS  & 2024 \\
      &                                 & Geo-MMRAG~\cite{chen2025vision}                      & Lithology Identification                           & Contrastive Trans.& CUG         & 2024 \\
    \cmidrule{2-7}
      & \multirow{7}{*}{Signal \& Geophysical Analysis}
        & Fish~\cite{zhang2024fast}                            & Earthquake Monitoring \& Early Warning             & RetNet          & PJLAB         & 2024 \\
      &                                 & SeisT~\cite{li2024seist}                             & Earthquake Monitoring                              & CNN+Transformer & CUMT          & 2024 \\
      &                                 & SeisCLIP~\cite{si2024seisclip}                       & Earthquake Class. \& Localization                  & Contrastive Trans.& USTC       & 2024 \\
      &                                 & InversionNet~\cite{wu2019inversionnet}               & Subsurface Imaging                                 & CNN             & Los Alamos    & 2019 \\
      &                                 & DispFormer~\cite{liu2025dispformer}                  & Subsurface Imaging                                 & Transformer     & SJTU          & 2025 \\ 
      &                                 & NCS-SFM~\cite{ordonez2026ncs}                        & Basin-scale Seismic Interpretation                 & ViT-MAE         & NR          & 2025 \\
      &                                 & TGS-SFM~\cite{sansal2025scaling}                     & Scalable Seismic Pretraining                       & ViT-MAE         & TGS           & 2025 \\
    \midrule

    \multirow{4}{*}{\shortstack[c]{Unified Earth FMs}}
      & \multirow{1}{*}{Textual Reasoning \& QA}
        & K2~\cite{deng2024k2}                                 & General Geoscience QA                              & Transformer     & SJTU          & 2025 \\
    \cmidrule{2-7}
      & \multirow{2}{*}{Multimodal Understanding}
        & TerraMind~\cite{jakubik2025terramind}                & Earth Observation Multimodality                    & Transformer     & IBM           & 2025 \\
      &                                 & PANGAEA GPT~\cite{pantiukhin2026hierarchical}        & Earth Science Multimodal QA                        & Transformer     & AWI   & 2026 \\
    \cmidrule{2-7}
      & \multirow{1}{*}{Agentic Workflows}
        & Earth-Agent~\cite{shabbir2026openearthagent}         & Autonomous Scientific Workflow                     & Agentic         & SHAI   & 2026 \\
    \bottomrule
    \end{tabular}
  }
  \label{tab:sphere_models_all} 
\end{table*}

\subsection{Biosphere Science}

Ecology sits a core of earth science because it links biophysical processes with human systems. As ecological change accelerates under global warming, land-use shifts, and urbanization, the field increasingly depends on integrative, data-rich, and computationally grounded approaches to monitor, understand, and govern complex socio-ecological dynamics. This section outlines the data foundations powering this shift, the principal task categories where AI and LFMs can add value, and the emerging landscape of LFMs for ecoscience. 

 \begin{figure*}[h]
	\centering
	\includegraphics[width=1.0\textwidth]{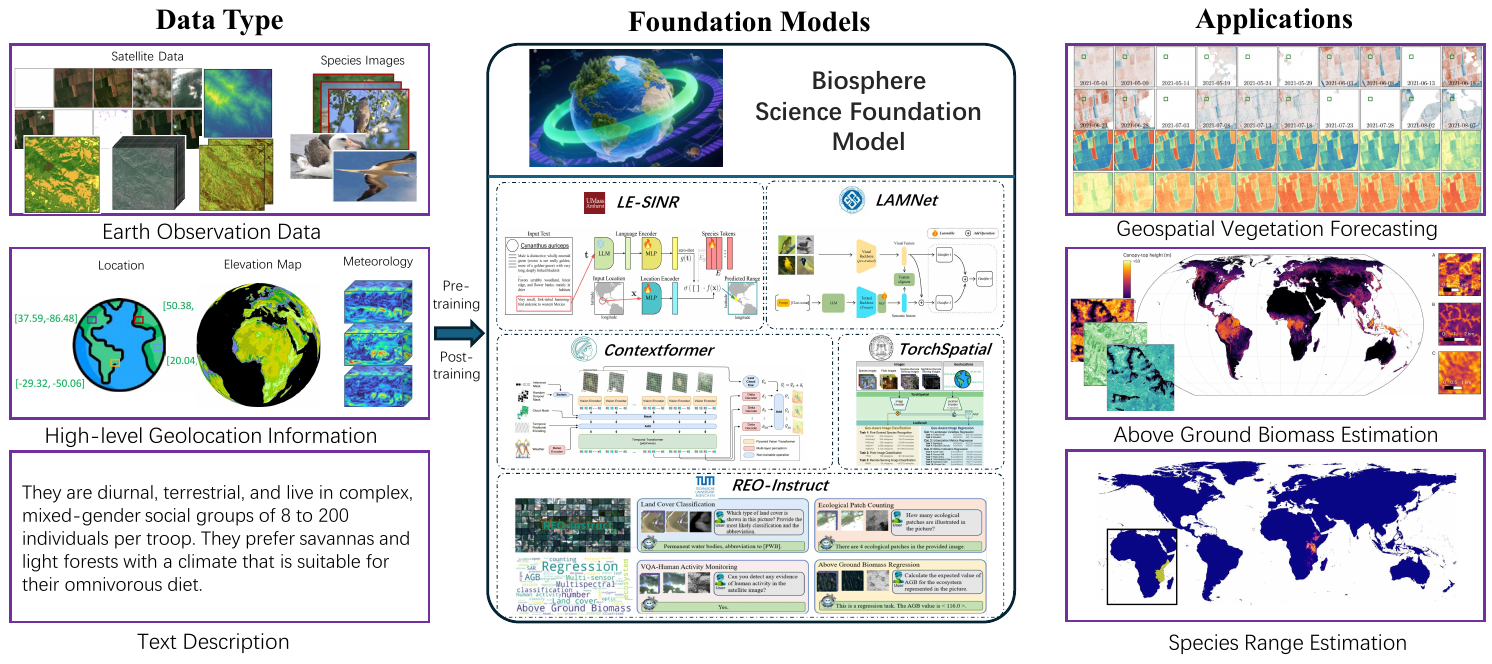}
        \caption{Illustration of biosphere science data type, foundation models and applications.
        }
        \label{bio_case}
\end{figure*}

\subsubsection{Data}

Ecological research increasingly relies on heterogeneous, large-scale data streams that capture both biophysical dynamics and human influences. Historically, trait-based ecology has conducted research based on the analysis of trait data collected at the individual level, including measurable morphological, physiological, phenological or behavioral characteristics, that impact fitness and provide a mechanistic understanding of ecological patterns ~\cite{de-bello-2021}. Nowadays, remote sensing technologies have enabled the capture of spatial distribution data and compilation of extensive satellite image datasets that, when integrated with global biodiversity databases (e.g., GBIF), facilitate comprehensive tracking of changes in land cover patterns, net primary productivity, and urban expansion ~\cite{morera-2024}. Contemporary ecological data streams encompass real-time information collected from sensor networks and Internet of Things (IoT) devices deployed across ecosystems ~\cite{morera-2024}. Complementing these technological sources is the expanding volume of data generated through social media platforms where volunteer observations and public discourse contribute ecological information ~\cite{morera-2024, doi-2023}. Furthermore, socio-economic datasets containing information on human populations, activities, and development trajectories provide essential context for understanding the complex interplay between anthropogenic factors and ecological systems ~\cite{morera-2024}. This multi-faceted data ecosystem offers rich opportunities for LFMs to extract patterns, generate insights, and support evidence-based decision-making in ecological research and conservation planning. The landscape of ecological research has been transformed by the proliferation of diverse datasets that, when leveraged through LFMs that can assimilate ecological multimodal data which is difficult in the traditional model, offer unprecedented analytical capabilities. 

\subsubsection{Tasks}

For biosphere science, we further dissect the existing literature into six primary classes based on their ecological objectives, which can be logically grouped into three overarching application domains. 
First, for \textbf{Species Distribution and Biodiversity}, the tasks include \textit{Data Processing and Trait Identification} and \textit{Species and Range Mapping}, which involve processing multimodal ecological data (e.g., morphology traits, occurrence records) to map distributions. 
Second, for \textbf{Ecosystem Monitoring and Risk Assessment}, the tasks comprise \textit{Environmental Monitoring} and \textit{Ecological Risk Estimation}, focusing on tracking ecosystem dynamics and forecasting environmental threats such as extreme events, biological invasions, or carbon budget shifts~\cite{morera-2024}. 
Finally, for \textbf{Ecological Decision Support and Agents}, the tasks encompass \textit{Decision Support and Automation} and \textit{Ecological Knowledge Consultation}, which leverage the reasoning and generative capabilities of LLMs to provide intelligent interfaces, automate analytical workflows, and assist researchers and policymakers in scientific programming and conservation management~\cite{morera-2024, dorm-2025, dsouza-2025}. 

\subsubsection{Applications}

The integration of Earth FMs into biosphere applications is reshaping how researchers and policymakers tackle complex ecological challenges. With their advanced multimodal understanding, reasoning, and generation capabilities, these models enable intelligent interaction with biological systems, providing solutions in areas such as biodiversity tracking, ecosystem monitoring, marine ecology, and environmental conservation~\cite{morera-2024, dorm-2025}. To illustrate this paradigm shift, the following sections highlight several pioneering works that demonstrate the transformative potential of Earth FMs across the three primary domains identified above.

\textbf{Species Distribution and Biodiversity:} Earth FMs offer transformative benefits to biodiversity monitoring by synthesizing textual habitat descriptions with spatial occurrence data. A prominent application is \textit{Species Range Estimation}. For instance, LE-SINR (Language Enhanced Spatial Implicit Neural Representation)~\cite{hamilton-2024} aligns a massive dataset of 35.5 million iNaturalist observations with Wikipedia text embeddings, enabling zero-shot species range map estimation from habitat descriptions and serving as a robust prior for few-shot learning. Identifying specific species in complex environments requires advanced trait perception; to this end, LAMNet introduces an LLM-assisted multi-modal learning framework for fine-grained few-shot image classification, significantly enhancing urban biodiversity monitoring. Enhancing the spatial fairness of these models relies on robust Spatial Representation Learning (SRL), as demonstrated by the TorchSpatial framework, which mitigates geographic bias in species recognition. In the realm of biological knowledge retrieval, BIORAG~\cite{wang-2024} introduces a Retrieval-Augmented Generation (RAG) framework that incorporates specialized biological databases (e.g., Gene, dbSNP) and iterative self-evaluation to achieve state-of-the-art performance in biological question answering. Furthermore, multi-agent frameworks like BioSim~\cite{dsouza-2025} explore biological invasions by enabling species agents to interact dynamically based on ecological principles and real-time literature.

\textbf{Ecosystem Monitoring and Risk Assessment:} Advanced foundation models are increasingly deployed to monitor ecosystem health, assess risks, and track structural changes from remote sensing imagery. A critical task here is \textit{Geospatial Vegetation Forecasting}, which models the temporal dynamics of plant health. To address this, Contextformer utilizes a lightweight multi-modal transformer to predict within-year vegetation greenness by fusing high-resolution Sentinel-2 satellite imagery with meteorological time series. Another key application is \textit{Above Ground Biomass (AGB) Estimation}. REO-Instruct serves as a pioneering unified benchmark that bridges qualitative visual descriptions with quantitative scientific regression, enabling Vision-Language Models (VLMs) to accurately perform AGB regression and ecological patch counting. Furthermore, Depth Any Canopy (DAC)~\cite{cambrin-2024} fine-tunes a monocular depth foundation model to estimate forest canopy height from aerial imagery, achieving superior accuracy compared to traditional remote sensing baselines. For post-disturbance recovery, HealingFactor~\cite{dsouza-2025} utilizes satellite imagery and deep learning to forecast ecosystem regeneration in war-affected regions by analyzing temporal vegetation patterns. Additionally, algorithms like EcoLogic~\cite{dsouza-2025} evaluate LFMs' ability to differentiate between causal and correlational reasoning within complex ecological systems, using food-web-based tasks from the Global Biotic Interactions (GloBi) dataset to assess ecosystem dynamics.

\textbf{Ecological Decision Support and Agents:} Intelligent systems driven by LLMs are revolutionizing environmental management by providing actionable insights to practitioners and policymakers. ChatEcology~\cite{yang-2025} exemplifies this in marine science; developed through domain-specific fine-tuning and expert validation, it functions as a conversational AI agent tailored for marine green tide research, achieving 89.4\% alignment with expert consensus. At the policy level, EcoGuard~\cite{dsouza-2025} operates as an AI-enhanced dashboard that integrates environmental datasets related to deforestation and carbon emissions to support sustainable governance. Similarly, specialized tools like Agri-Chatbot provide science-backed agricultural guidance via RAG, while EcoSearch optimizes literature retrieval based on geographical author distribution to foster a more inclusive scientific discourse~\cite{dsouza-2025}.

\subsection{Anthroposphere Science}

 \begin{figure*}[h]
	\centering
	\includegraphics[width=1.0\textwidth]{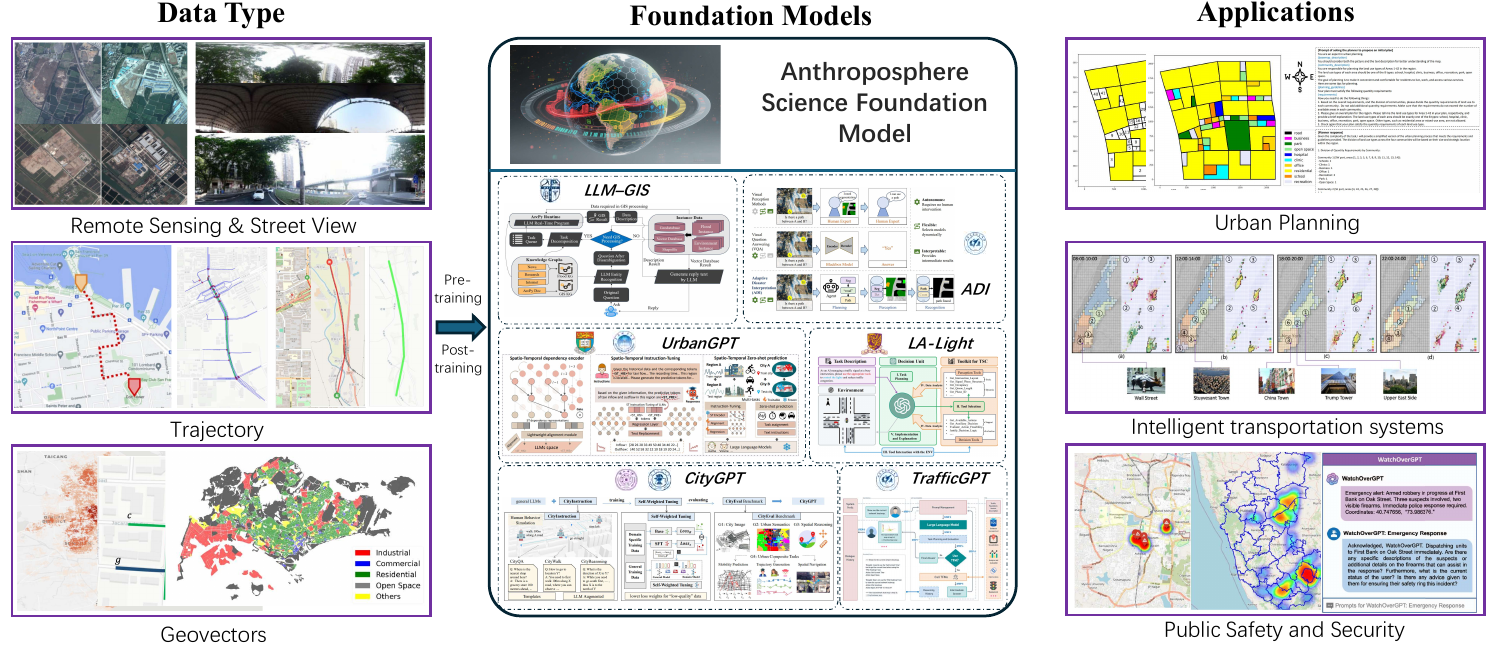}
        \caption{Illustration of anthroposphere science data type, foundation models and applications.
        }
        \label{anthro_case}
\end{figure*}

Earth science foundation models play a crucial role in comprehending and managing Earth’s complex systems and environments, serving as a formidable toolset to glean insights, anticipate trends, and devise effective strategies for sustainable development and resource management. They are becoming foundational in the assessment of urban development, enabling comprehensive analyses across diverse aspects such as urban expansion and infrastructure changes. By leveraging MLLM-powered algorithms, urban planners and policymakers can evaluate the spatial dynamics of urban growth, anticipate infrastructure needs, optimize transportation networks, and devise sustainable land-use strategies. By integrating geospatial data with advanced analytical techniques, MLLMs empower stakeholders to make informed decisions that promote resilient, inclusive, and environmentally sustainable urban environments.

\subsubsection{Data}

In urban studies and anthroposphere modeling, multi-modal dataset integration is essential for capturing complex human-environment dynamics. Geovector data (e.g., points, lines, and polygons) serves as the fundamental input for spatial representation learning~\cite{mai2022review}. Furthermore, large-scale human mobility data, derived from aggregated mobile location services and GPS traces, provides critical context for macro-scale interactions~\cite{yabe2022mobile}. Rather than merely supporting traffic routing, these modalities record population movements under environmental stress. This empowers Earth scientists to model mass displacements during natural disasters~\cite{lu2012predictability}, evaluate dynamic hazard exposure, and forecast human flow disruptions during extreme weather~\cite{chen2020impact}, thereby directly supporting planetary health modeling.

Crucially, visual data collected from satellites and street-level cameras provides the rich perceptual information necessary to monitor the physical anthroposphere. The recent proliferation of remote sensing (RS) image-text pairs has driven massive advancements in vision-language pretraining. Datasets such as RS5M (GeoRSCLIP)~\cite{zhang2024rs5m}, RemoteCLIP~\cite{liu2024remoteclip}, SkyCLIP~\cite{wang2024skyscript}, S-CLIP~\cite{mo2023s}, and RS-CLIP~\cite{li2023rs} enable robust cross-modal representation learning and zero-shot classification. These capabilities are vital for rapidly mapping urban sprawl or assessing infrastructure damage post-disaster. Additionally, geo-tagged multispectral datasets like S2-100K~\cite{klemmer2025satclip} provide critical temporal (2010--present) and spatial context, anchoring visual observations to precise geo-locations to track long-term urban evolution.

To bridge the gap between raw visual perception and high-level semantic reasoning, researchers rely heavily on instruction-tuning and visual QA datasets. Resources such as SkyEye-968k~\cite{zhan2025skyeyegpt}, RSICap~\cite{hu2025rsgpt}, MMRS-1M~\cite{zhang2024earthgpt}, LHRS-Align/Instruct~\cite{muhtar2024lhrs}, and RSIVQA~\cite{lobry2020rsvqa} equip foundation models with the ability to interpret expert-annotated captions and execute complex reasoning tasks. Furthermore, these visual and spatial capabilities must be combined with textual data derived from policy documents, regulatory codes, and social media. This final layer of integration ensures comprehensive spatial, temporal, and semantic coverage, enabling AI models to align physical urban changes with societal priorities and governance frameworks.

\subsubsection{Tasks}

We categorize the existing literature into several key domains of application within urban environments: Transportation, Urban Planning, Energy Management, Environmental Monitoring, Public Safety, and Disaster Mobility Modeling. In transportation, models process heterogeneous inputs (time series, trajectories, images, text) to enhance traffic management and transit routing. Urban planning tasks leverage large datasets to generate layout proposals and parse public feedback for participatory planning. Energy management focuses on optimizing domain-specific grid data while employing privacy-preserving techniques (e.g., federated learning) to safeguard sensitive infrastructure information. Environmental monitoring utilizes satellite telemetry to track climate variations and pollution. Public safety tasks focus on analyzing crime reports and integrating external tools for optimal emergency resource allocation. Concurrently, disaster mobility modeling utilizes trajectory and spatial data to simulate mass population displacements, evaluate dynamic exposure to environmental hazards, and coordinate evacuation strategies under extreme weather conditions.

\subsubsection{Applications}

The integration of Earth FMs into urban applications is fundamentally reshaping how cities tackle complex, multi-faceted challenges. With advanced multimodal understanding, reasoning, and generative capabilities, Earth FMs enable intelligent interaction with urban systems. As research rapidly progresses, reviewing these applications reveals critical trends in planning, transportation, public safety, environmental sustainability, and disaster resilience.

\textbf{Urban Planning:} Earth FMs offer transformative benefits to urban planning by analyzing extensive urban data and harnessing public feedback to optimize spatial layouts. 

On one hand, centralized urban planning leverages government-led initiatives to enhance infrastructure and management. Earth FMs facilitate this by parsing vast regulatory knowledge bases. For instance, UrbanPlanBench~\cite{zheng2025urbanplanbench} investigates LLMs' proficiency in acquiring planning principles and regulations. PlanGPT-VL~\cite{zhu2025plangpt} utilizes domain-specific fine-tuning to align with governmental document styles, enhancing the generation and evaluation of planning documents through tailored embedding models. Furthermore, agentic systems like City-LEO~\cite{jiao2024city} synergize LLM reasoning abilities with strict urban optimizers to derive robust decisions in highly uncertain environments.

On the other hand, participatory urban planning involves diverse stakeholders. FMs facilitate collaborative urban development by processing unstructured civic input. Zhou et al.~\cite{zhou2024large} propose a participatory land-use framework that mines public feedback from digital platforms. Singla et al.~\cite{singla2024adaptive} deploy specialized LLM agents to manage sub-area development by balancing local and regional land-use requirements. Additionally, Ni et al.~\cite{ni2024planning} introduce a cyclical planning paradigm that utilizes simulations and automated resident interviews for continuous plan evaluation and urban regeneration.

\textbf{Transportation:} Intelligent transportation systems (ITS) have revolutionized road safety~\cite{wang2023transportation} and transit routing~\cite{liu2020polestar}. Modern ITS now incorporate advanced predictive analytics enabled by LLMs. 

LLMs have emerged as critical tools for predictive analytics by integrating unstructured data (news reports, geospatial texts) into traditional forecasting models~\cite{ning2023uukg, huang2024enhancing, nie2025joint}. Methods like Open-TI~\cite{da2024open} support advanced simulations for demand optimization, while conversational systems like TransitGPT~\cite{devunuri2025transitgpt}, TrafficGPT~\cite{zhang2024trafficgpt}, and CityGPT~\cite{guan2024citygpt} allow users to query massive transportation datasets via natural language, democratizing access to mobility insights.

In the realm of traffic management, LLMs dynamically generate adaptive control strategies. Early work demonstrated LLMs acting as decision-making agents using chain-of-thought reasoning for signal control~\cite{lai2025llmlight}. Hybrid frameworks like LA-Light~\cite{wang2024llm} integrate LLMs with specialized tools to handle rare events like sensor failures, while iLLM-TSC~\cite{pang2407illm} supervises reinforcement learning policies to manage signals under imperfect observation. Cooperative multiagent systems (e.g., CoLLMLight~\cite{yuan2025collmlight}) take this further by constructing spatio-temporal graphs to predict optimal, city-wide signal configurations.

Furthermore, recent studies fuse LLMs directly with spatio-temporal neural networks. Models like ST-LLM~\cite{liu2024st} and TPLLM~\cite{ren2024tpllm} leverage fine-tuned attention to address missing data imputation and noise reduction, whereas UrbanGPT~\cite{li2024urbangpt} enriches predictions in data-scarce environments via context-aware prompting. This fusion paves the way for responsive, data-driven ITS capable of predicting event-driven traffic variations.

\textbf{Public Safety and Security:} Public safety represents a fundamental pillar of urban resilience. Recent research leverages advanced analytics to enhance crime prevention, emergency response, and disaster management. Foundation models utilize real-time data and multimodal inputs, such as satellite imagery and behavioral statistics, to predict and mitigate urban risks~\cite{chen2016learning,najjar2017combining,wickramasekara2025exploring,shahid2024watchovergpt}. Specialized models, including TrafficSafetyGPT~\cite{zheng2023trafficsafetygpt} and ChatLaw~\cite{cui2023chatlaw}, demonstrate the efficacy of instruction-tuned LLMs in generating reliable outputs in the traffic safety and legal domains. Researchers increasingly use LLMs as text encoders integrated with machine learning algorithms to enhance zero-shot crime prediction and accident severity classification~\cite{grigorev2025enhancing,sarzaeim2024experimental}. They also serve as intelligent assistants to automate accident reports and facilitate real-time surveillance~\cite{zheng2023chatgpt,shahid2024watchovergpt,mumtarin2023large,zhen2024leveraging,hostetter2024role,zhou2024gpt}. By integrating LLMs with external knowledge graphs and search engines, these systems drastically improve resource allocation and coordination~\cite{chen2024enhancing,de2023llm}.

\textbf{Disaster Resilience and Climate Mobility:} Expanding beyond routine urban management, Earth FMs and advanced simulation frameworks are increasingly deployed to model human-environment interactions under severe ecological stress. Understanding how populations move and respond to extreme events is critical for planetary health and emergency planning. Recent advancements address this by leveraging large language models (LLMs) to generate highly realistic, nuanced human mobility patterns. For example, generative frameworks like MobilityGPT enhance human trajectory modeling by utilizing GPT architectures to capture the complex spatial-temporal dynamics of population movements~\cite{haydari2025mobilitygpt}. Furthermore, LLM-empowered agent-based modeling (ABM) has emerged as a powerful paradigm for simulating complex societal behaviors in virtual sandbox environments~\cite{gao2024large}. By configuring LLMs to act as virtual urban residents, researchers can generate highly personalized mobility profiles and simulate individual decision-making processes under varying environmental contexts~\cite{wang2024large}. Integrating these sophisticated mobility and agent-based frameworks with real-time hazard data enables policymakers to dynamically anticipate population shifts, test response protocols, and ultimately ensure that vulnerable populations are efficiently protected during climate-induced crises.

\textbf{Environmental Sustainability:} Rapid urbanization generates severe environmental challenges, such as air pollution, resource depletion, and climate change, which threaten long-term urban viability~\cite{ullah2020applications}. Advanced monitoring systems leverage heterogeneous sensor networks collecting meteorological data, water quality metrics, and satellite land-cover imagery~\cite{yuan2020deep}. 

Deep learning architectures, such as DeepAir~\cite{zhao2023deepair} and AirFormer~\cite{liang2023airformer}, were developed to integrate these datasets, leading to substantially improved accuracy in forecasting air quality indices. Building on this, LLMs have been integrated as sophisticated virtual assistants. They address complex environmental queries, act as reasoning engines when combined with Geographic Information Systems (GIS) to improve flood risk perception~\cite{zhu2024flood}, and function as autonomous agents that synthesize real-time data into actionable mitigation insights~\cite{kraus2023enhancing}. 

In parallel, urban energy management and pollution control are benefiting from intelligent optimization. Techniques like LLM4DistReconfig~\cite{christou2025llm4distreconfig} optimize power grid configurations in near-real time. Other approaches improve waste management operations by simulating citizen perceptions and localized ecological impacts~\cite{verma2023generative}. Finally, because energy and grid data are highly sensitive, the adoption of privacy-preserving paradigms (e.g., federated learning) is heavily explored to securely scale these LLM-driven sustainability frameworks~\cite{yang2019federated}.

\subsection{Lithosphere Science}

 \begin{figure*}[h]
	\centering
	\includegraphics[width=1.0\textwidth]{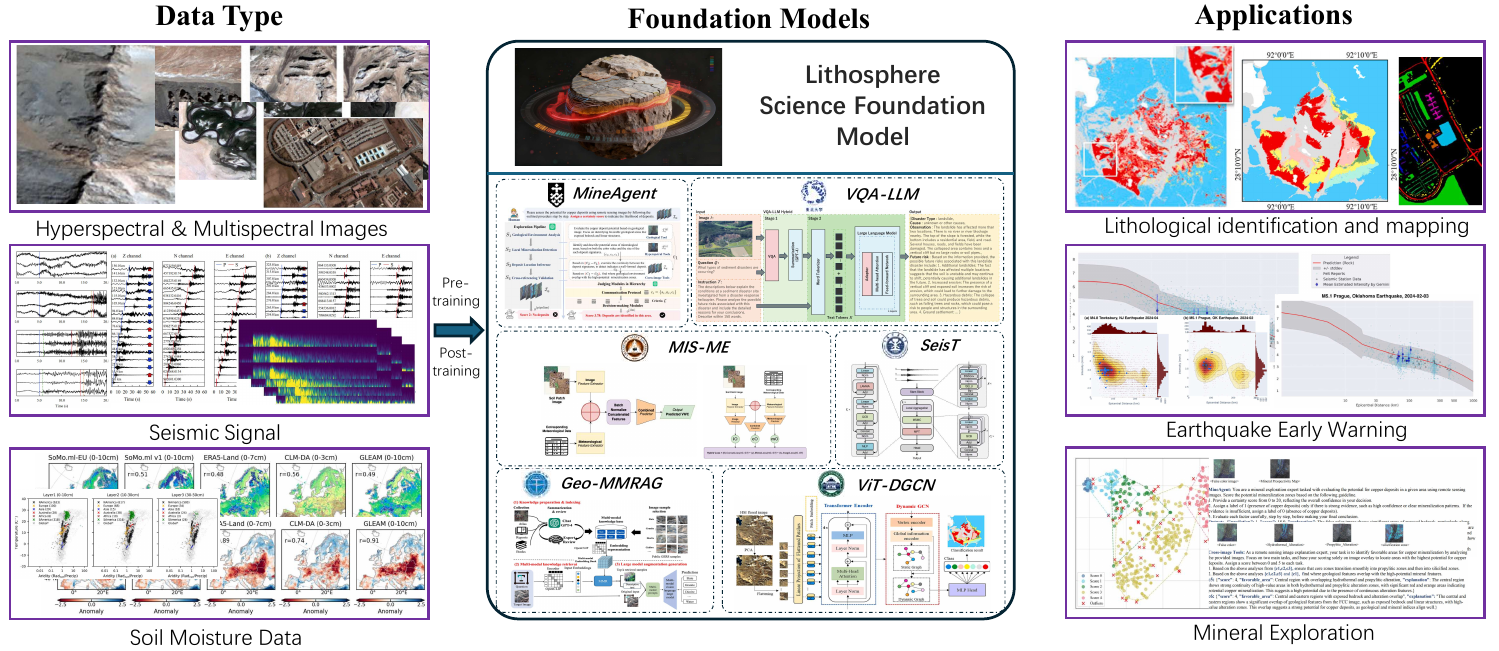}
        \caption{Illustration of lithosphere science data type, foundation models and applications.
        }
        \label{lith_case}
\end{figure*}

Lithosphere science, which focuses on understanding the structure, composition, and dynamic evolution of the Earth’s interior, stands to benefit immensely from recent advances in AI and foundation models. These models provide a powerful framework for integrating multi-modal geological data, constructing comprehensive subsurface representations, and identifying patterns associated with lithospheric processes. By integrating diverse data sources—including geological, geophysical, and geochemical measurements—Earth science foundation models can enhance subsurface imaging and tectonic interpretation, thereby supporting more precise monitoring and predictive analyses of geological hazards such as earthquakes and landslides~\cite{zhang2023marine,areerob2025multimodal}. Furthermore, AI-driven techniques are revolutionizing lithological classification, rock property inversion, and mineral resource exploration~\cite{liu2024foundation}. Overall, the integration of AI and large-scale data analytics in lithosphere science is paving the way for innovative approaches in geological hazard prediction, resource exploration, and environmental management, advancing our understanding of Earth’s lithospheric dynamics and enabling a safer and more sustainable use of geological resources~\cite{zhao2024artificial}.

\subsubsection{Datasets}

The application of AI in lithosphere science leverages a diverse array of datasets that capture the multifaceted nature of the Earth's subsurface. These heterogeneous and multi-modal datasets underpin Earth science foundation models designed for lithospheric studies. In practice, geoscience data include high-resolution remote sensing imagery (e.g., hyperspectral or multispectral images capturing rock textures and mineralogical patterns), structured geochemical measurements, geophysical survey records (e.g., active and passive seismic recordings, gravity and magnetic observations), well-log data, and unstructured text from geological reports. Unimodal datasets—such as raw field images and spectral profiles—can be enhanced by cross-modal information extracted from digital reports via natural language processing tools (e.g., the GeoDocA system~\cite{holden2019geodoca}). In addition, the integration of such diverse data sources enables Earth science foundation models to learn comprehensive geological representations and to mitigate the challenges of high-dimensional feature spaces encountered in mineral mapping~\cite{lorenz2021feature}.

\subsubsection{Tasks}

In lithosphere science, AI-driven research addresses several key tasks, including geological exploration and hazard prediction by integrating and analyzing heterogeneous geophysical datasets (including seismic, well-log, magnetotelluric, and gravity measurements) to enhance the detection of subsurface reservoirs and natural hazards~\cite{yu2021deep}. Additionally, AI facilitates rock physics analysis through automated classification and interpretation of rock samples. It also enhances geological modeling by combining multidisciplinary data ranging from geophysics to tectonics to reconstruct complex subsurface structures. Finally, AI advances deep-time and deep-Earth investigations by synthesizing observational and synthetic data to unravel intricate lithospheric processes~\cite{wang2021deep}.

\subsubsection{Applications}

Earth FMs have demonstrated significant potential in transforming traditional workflows in lithosphere research by enabling multi-task learning and robust feature fusion across different data modalities. The following examples illustrate specific applications incorporating detailed model implementations and references from recent studies.

\textbf{Mineral Exploration:}
Recent advances in mineral exploration increasingly leverage machine learning to assist geoscientists in streamlining data analysis and interpretation, spanning tasks from regional targeting to resource delineation~\cite{woodhead2021harnessing,zuo2024explainable}. Earlier ML applications primarily focused on classification models applied to diverse geoscientific datasets~\cite{taheri2021long}, including lithogeochemical, geophysical, remote sensing, and textual data. However, despite the availability of multiparameter datasets, these datasets rarely meet the 'big data' criteria as defined by De Mauro et al.~\cite{de2015big}, and conventional ML approaches have face challenges in effectively integrating complex spatial, spectral, and geological context.
To address these limitations, recent efforts have turned to multimodal large language models (MLLMs), which offer novel capabilities for simultaneous integration of heterogeneous data sources. For instance, MineAgent~\cite{yu2024mineagent} exemplifies how MLLMs can be adapted for remote sensing-based mineral exploration by effectively handling multiple remote sensing images and large text corpora concurrently. MineAgent employs hierarchical judgment and decision-making modules to extract, integrate, and analyze spatial and spectral features from hyperspectral imagery, while also incorporating domain-specific geological terminology. By orchestrating information from as many as nine images, MineAgent addresses the long-context issues that previously hindered the accurate synthesis of spatial relationships across geological contexts.

\textbf{Hazard Prediction (e.g., Earthquake Detection and Forecasting):}
MLLM have been increasingly employed to enhance earthquake monitoring and forecasting. For instance, SeisT~\cite{li2024seist} integrates three-component seismic waveforms with additional contextual seismic information using a hybrid architecture that combines convolutional layers with Transformer-based self-attention mechanisms. SeisT has been successfully applied to multiple tasks, including seismic event detection, phase picking, and magnitude estimation, thereby providing improved real-time monitoring and early warning capabilities. Similarly, the SeisCLIP model~\cite{si2024seisclip} utilizes contrastive learning on seismic spectrograms and event metadata to support event classification and localization, further reducing prediction uncertainties in fault zone dynamics. Furthermore, SeisMoLLM~\cite{wang2025seismollm} applies cross-modal transfer learning from pretrained large language models to achieve state-of-the-art performance across diverse seismic monitoring tasks with high efficiency and strong generalization.

\textbf{Rock Physics Analysis:}
In rock physics analysis, MLLM foundation models leverage large-scale image and spectroscopic datasets to automatically classify rock samples and extract detailed textural and compositional features. For instance, Dong et al.~\cite{dong2024fusion} introduced a hybrid architecture combining vision Transformers with graph convolutional networks to learn both geometric and spectral features directly from rock images. This approach enables the precise differentiation among rock types and provided more refined input parameters for rock mechanics models, thereby advancing the understanding of deep-earth processes beyond traditional empirical methods.

\textbf{Geological Modeling:}
Deep learning methods play a crucial role in constructing high-fidelity subsurface models by integrating multi-modal data from geophysical surveys, geochemical analyses, and hydrological investigations. For instance, models that combine multilayer perceptrons with convolutional neural networks have been successfully employed to fuse surface and subsurface data, generating detailed predictive models of lithological interfaces, stratigraphic configurations, and groundwater flow paths~\cite{chen2020three}. Moreover, Guo et al.~\cite{guo2025deep} and Liu et al.~\cite{liu2025deep} leverage deep reparameterization to simultaneously recover multiple subsurface properties (e.g., velocity, gravity, and electromagnetic properties) from multi-modal datasets, effectively bridging data-driven and physics-constrained modeling. Such integrative approaches reduce the need for labor-intensive human-computer interactions and substantially enhance model robustness, even in data-scare regions.


In summary, from the GeoDocA and hybrid exploration models of Wang et al. (2021a) to the seismic monitoring implementations of SeisT (Li et al., 2024) and SeisCLIP (Si et al., 2023), as well as the advanced rock physics (Dong et al., 2024) and geological modeling (Chen et al., 2020; Jia et al., 2021), MLLM foundation models are rapidly evolving the landscape of lithosphere research. These methods not only streamline the data processing pipeline and mitigate traditional methodological limitations but also substantially enhance the efficiency, accuracy, and scalability of mineral resource exploration and geological hazard prediction.

\subsection{Hydrosphere Science}

Hydrosphere science studies the distribution, dynamics, and interactions of water across oceans, rivers, lakes, wetlands, and coastal environments. Recent advances in AI and foundation models provide new tools for integrating heterogeneous hydrosphere data, including remote sensing imagery, underwater videos, environmental variables, and textual knowledge. These methods support a wide range of applications, such as flood monitoring, waterbody mapping, marine ecosystem observation, and maritime decision-making~\cite{zheng2023marinegptunlockingsecretsocean,bi-etal-2024-oceangptacl,ren2024watergpt}.

\subsubsection{Datasets}
Hydrosphere datasets are highly diverse and often overlap with broader geoscience resources. Common data types include aerial and satellite imagery, underwater photographs and videos, hydrological variables, and multimodal image-text pairs. For flood-related studies, FloodNet~\cite{FloodNet} is a representative dataset containing high-resolution post-disaster aerial imagery collected after Hurricane Harvey. It provides labels for flood scenes, semantic segmentation masks, and visual question answering (VQA) pairs, supporting multimodal flood assessment. For marine science, the J-EDI dataset~\cite{yoshida2024jediqabenchmarkdeepsea} contains deep-sea images and videos collected during JAMSTEC submersible missions, together with metadata and annotated subsets for organism detection. Large-scale image-text corpora have also been developed for marine vision-language learning, such as the Marine-5M dataset used in MarineGPT~\cite{zheng2023marinegptunlockingsecretsocean}. Habitat-specific datasets are also important. CoralMask~\cite{CoralSCOP}, introduced with CoralSCOP, provides 41,297 images and 330,144 coral masks for dense coral segmentation. Fine-grained coral datasets with taxonomy labels and morphology descriptions further support detailed coral recognition~\cite{enhancingHan_Wang_Zhang_Li_Wang_2025}.

\subsubsection{Tasks}

Hydrosphere AI tasks mainly include image classification, semantic and instance segmentation, object detection, spatiotemporal prediction, and vision-language reasoning. These tasks are used to identify flooded regions, extract water bodies, detect marine organisms, segment coral habitats, and answer scientific or operational questions about aquatic environments~\cite{FloodNet,yoshida2024jediqabenchmarkdeepsea,CoralSCOP}.

\subsubsection{Applications}

Current applications of foundation models in hydrosphere science can be grouped into several directions.

\textbf{Flood monitoring and waterbody mapping.}
AI models are increasingly used for flood extent mapping, disaster assessment, and surface water extraction. FloodNet~\cite{FloodNet} supports flood classification, segmentation, and question answering in post-disaster scenarios. For remote sensing waterbody extraction, MF-SegFormer improves the delineation of rivers and surface water boundaries from Landsat imagery~\cite{WaterBody}.

\textbf{Marine ecological monitoring.}
Marine ecosystems are challenging because of low visibility, sparse annotations, and long-tail category distributions. MarineGPT~\cite{zheng2023marinegptunlockingsecretsocean} uses large-scale marine image-text pairs to improve marine object recognition and scientific response generation. MarineDet~\cite{haixin2023marinedetopenmarineobjectdetection} addresses open-marine object detection across 821 categories, while MarineInst~\cite{MarineInst} extends this line to instance-level masks and captions for marine objects.

\textbf{Coral reef analysis.}
CoralSCOP~\cite{CoralSCOP} provides a foundation model for dense coral segmentation and shows strong zero-shot generalization across reef environments. CORAL-Adapter~\cite{enhancingHan_Wang_Zhang_Li_Wang_2025} further improves fine-grained coral recognition by injecting coral taxonomy and morphology knowledge into CLIP-based models.

\textbf{Water science reasoning and maritime intelligence.}
OceanGPT~\cite{bi-etal-2024-oceangptacl} is designed for ocean science question answering and instruction following. WaterGPT~\cite{ren2024watergpt} targets multimodal hydrology tasks through a multi-agent framework. LITE~\cite{li2024lite} models environmental systems by converting heterogeneous variables into language and line-graph representations. For maritime applications, KUNPENG~\cite{wang2024kunpengembodiedlargemodel} integrates multi-source environmental data for intelligent maritime decision-making, and Llamarine~\cite{nguyen2025llamarineopensourcemaritimeindustryspecific} supports navigation-related tasks such as trajectory planning and risk assessment.

 \begin{figure*}[h]
	\centering
	\includegraphics[width=1.0\textwidth]{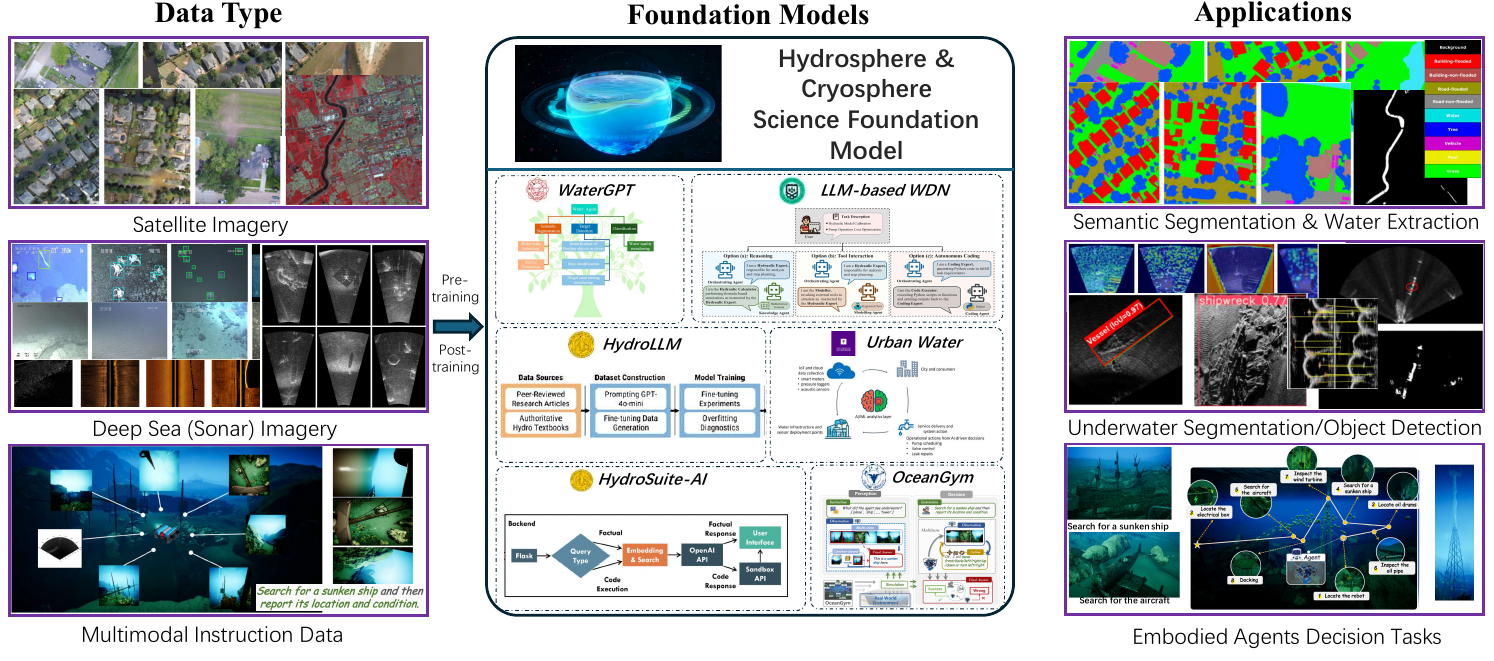}
        \caption{Illustration of hydrosphere\&cryosphere science data type, foundation models and applications.
        }
        \label{hydro_case}
\end{figure*}

\subsection{Cryosphere Science}

Cryosphere science focuses on the frozen components of the Earth system, including sea ice, glaciers, ice sheets, snow cover, and permafrost. These components strongly influence climate regulation, sea-level change, freshwater resources, and polar ecosystems. AI and foundation models are increasingly used in cryosphere science to learn spatiotemporal patterns from remote sensing observations and geophysical records, with particular progress in sea ice monitoring and forecasting~\cite{finn2024diffusionmodelslargescaleseaice,IceDiffXu_2025_CVPR,Spatiotemporal10402090,SICFormerjmse12081424}.

\subsubsection{Datasets}

Cryosphere datasets mainly come from satellites, airborne measurements, in situ observations, and reanalysis products. Common data types include passive microwave observations, SAR and optical imagery, sea ice concentration and thickness products, snow cover maps, and meteorological variables. These datasets are typically large-scale spatiotemporal grids with strong seasonality and uncertainty, making them suitable for transformer-based and diffusion-based models~\cite{finn2024diffusionmodelslargescaleseaice,IceDiffXu_2025_CVPR}.

\subsubsection{Tasks}

The main AI tasks in cryosphere science include observational mapping, spatiotemporal forecasting, and generative simulation. Typical examples include sea ice concentration prediction, ice thickness estimation, and generation of realistic future ice states. Compared with static image understanding, cryosphere tasks emphasize long-range temporal dependencies and uncertainty-aware prediction~\cite{Spatiotemporal10402090,SICFormerjmse12081424,finn2024diffusionmodelslargescaleseaice}.

\subsubsection{Applications}

\textbf{Sea ice forecasting.}
Sea ice prediction is currently the most active AI application in cryosphere science. IceFormer~\cite{Spatiotemporal10402090} combines empirical orthogonal function decomposition with transformer-based modeling to predict daily Arctic sea ice concentration and thickness up to 45 days ahead. SICFormer~\cite{SICFormerjmse12081424} uses a 3D Swin Transformer encoder and PixelShuffle decoder for short-term pan-Arctic sea ice concentration forecasting.

\textbf{Generative modeling of sea ice.}
Generative models are increasingly used to produce realistic ensembles of future sea ice states. Finn et al.~\cite{finn2024diffusionmodelslargescaleseaice} develop latent diffusion models for Arctic-wide sea ice generation with physical bounds, achieving much higher computational efficiency than data-space diffusion models. IceDiff~\cite{IceDiffXu_2025_CVPR} combines a vision transformer forecasting model with diffusion-based generation to produce high-resolution sea ice concentration predictions at 6.25 km resolution.

\textbf{Climate monitoring and polar risk assessment.}
These models are also useful for broader climate and operational applications, including polar navigation, environmental risk assessment, and uncertainty-aware monitoring of Arctic change~\cite{finn2024diffusionmodelslargescaleseaice,IceDiffXu_2025_CVPR,Spatiotemporal10402090}. Although current work is concentrated on sea ice, similar approaches are likely to expand to glacier, snow, and permafrost studies in the future.

%% file: draft/040_discussion.tex
\input{draft/041_extensions}

\input{draft/042_challenges}

%% file: draft/041_extensions.tex
\section{Unified Earth FMs}
\label{sec:extensions}

With the rapid advancement of LLM/MLLMs, earth science research is undergoing a paradigm shift from single-domain, task-specific modeling to integrated frameworks capable of addressing the Earth system's inherent complexity. Traditional earth science foundation models have long been constrained by modality silos, shallow reasoning capabilities, and limited integration across Earth's spheres. In contrast, unified Earth science foundation models are emerging as a transformative force, characterized by four core pillars: multi-sphere integration; advanced reasoning enabling complex logical deduction with sparse supervision; agentic collaboration through tool-augmented workflows; and accelerated scientific discovery via knowledge synthesis and pattern recognition. This chapter systematically reviews these developments, integrating foundational advances and cutting-edge research to delineate the evolving landscape of unified earth science AI.

\subsection{Multi-Sphere Fusion and Integration}

\subsubsection{Rationale and Challenges}

The Earth system's interconnectedness, where atmospheric circulation drives ocean currents, terrestrial vegetation modulates carbon cycles, and subsurface processes influence surface geology, demands models that transcend traditional disciplinary divides. Historical Earth science models achieved limited integration through physical process coupling but struggled with data heterogeneity and computational scalability. Modern MLLM-based frameworks address these gaps by unifying diverse data modalities across Earth's spheres, leveraging self-supervised learning to extract common representations from unstructured, multi-source data.
Key challenges in multi-sphere integration include: (1) sensor heterogeneity across missions (e.g., SAR, hyperspectral, thermal infrared); (2) spatial-temporal scale mismatches (e.g., hourly atmospheric measurements vs. annual geological changes); (3) sparse supervision in extreme environments (e.g., polar regions, deep oceans); and (4) metadata integration (e.g., geospatial coordinates, sensor parameters, temporal stamps) that enriches semantic understanding beyond raw imagery.

\subsubsection{State-of-the-Art Frameworks}

Copernicus-FM~\cite{wang2025towards} represents a milestone in multi-sphere integration, unifying 18.7M aligned observations from all major Copernicus Sentinel missions (Sentinel-1 to 5P) to bridge surface and atmospheric monitoring. This enables seamless transitions between tasks like land cover classification (surface) and air quality estimation (atmosphere), with joint pretraining improving performance across both domains by 12-18\% compared to single-sphere models.
Complementary to Copernicus-FM, ESA and IBM's TerraMind~\cite{jakubik2025terramind} model integrates 8 data types (including Sentinel imagery, terrain, and land use data) through self-supervised learning, achieving state-of-the-art performance in cross-sphere tasks like methane leak detection (atmospheric) and forest cover change tracking (terrestrial). Its generative capabilities further address data sparsity by synthesizing missing modalities, a critical advantage for polar and remote regions where observations are limited.

SkySense++~\cite{guo2024skysense} advances multi-sphere integration through a two-stage pretraining paradigm: first capturing cross-modal, multi-granular (pixel-object-image) representations, then enhancing semantic understanding via masked semantic learning. Trained on 2700 data from 11 satellite, it achieves 4.79\% higher average precision in full-fine-tuning tasks and 14.08\% higher mIoU in few-shot segmentation compared to prior models. Notably, its "prompt-only" deployment capability eliminates the need for task-specific fine-tuning, enabling rapid adaptation to new multi-sphere scenarios like agricultural drought monitoring (integrating vegetation indices and precipitation data) and volcanic hazard assessment (combining thermal imagery and seismic metadata).

\subsection{Unified Reasoning and Deduction}

\subsubsection{Limitations of Traditional Approaches}

Conventional geospatial Vision-Language Models rely heavily on Supervised Fine-Tuning and contrastive learning, which often fail in complex Earth Observation settings due to brittle in-domain stability, poor out-of-distribution generalization, and the catastrophic forgetting of general capabilities. Furthermore, models trained via standard SFT tend to exhibit "shallow reasoning"—memorizing textual priors rather than developing structured chain-of-thought processes. This stems from the sparse, metadata-only supervision typical in geospatial datasets, which lacks the dense, verifiable reasoning traces required for tasks like multi-temporal change detection or cross-view localization.

\subsubsection{Reinforcement Learning for Reasoning}

To overcome this SFT data bottleneck, recent research has shifted towards post-training frameworks that leverage Reinforcement Learning with verifiable rewards (e.g., RLVR, GRPO). Unlike static supervised learning, RL permits models to explore reasoning paths and receive feedback based on outcome correctness, transitioning them from basic perception engines to robust deductive systems. Because obtaining detailed reasoning traces is prohibitively expensive in the geospatial domain, several distinct methodologies have emerged to formulate effective reward signals. For instance, approaches like Geo-R1~\cite{zhang2026geo} utilize weakly-supervised proxies, formulating cross-view pairing tasks (e.g., matching street-level imagery to satellite counterparts) to enhance multi-modal synthesis and tiny visual cue extraction without manual reasoning labels. To handle the inherent diversity of EO tasks, frameworks such as GeoVLM-R1~\cite{fiaz2025geovlm} introduce task-aware dual-objective rewards that combine spatial precision metrics (such as bounding-box IoU) with semantic alignment scores. Furthermore, to combat the generation of unverifiable or logically drifting outputs, models like GeoReason~\cite{li2026georeason} and RSThinker~\cite{liu2025towards} emphasize "faithful reasoning" by penalizing logical drift and enforcing strict alignment between visual evidence and final conclusions.



\subsection{Agentic Workflows for Discovery}

\subsubsection{From Perception-Centric Models to Executable Scientific Workflows}

Recent Earth AI systems increasingly extend beyond perception-oriented tasks such as scene understanding, land-cover classification, and object detection. In many real-world settings, Earth science analysis requires not only interpreting observations, but also retrieving data, invoking domain tools, executing code, and refining intermediate results over multiple steps. This creates a gap between conventional MLLMs, which are primarily optimized for single-turn multimodal understanding, and the more procedural workflows common in geoscientific practice.

This gap has motivated growing interest in \emph{agentic frameworks}, where language models are coupled with external tools, memory, and executable environments. Rather than producing one-shot answers, such systems aim to plan and carry out analysis workflows that better match how Earth scientists interact with heterogeneous data, software libraries, and scientific repositories. 

\subsubsection{Tool-Augmented Agent Systems}

The first major direction in this transition is the development of tool-augmented single-agent systems, which couple an LLM reasoning core with external tools for data access, code execution, and scientific computing. These systems extend the capabilities of pretrained models by allowing them to call domain-specific libraries, query remote repositories, and execute analysis steps in a persistent environment.

Recent frameworks demonstrate the effectiveness of this design. \textit{GISclaw}~\cite{han2026gisclaw}, for example, integrates an LLM with a persistent Python sandbox and a comprehensive GIS software stack, including GeoPandas, rasterio, scipy, and scikit-learn, enabling full-stack geospatial analysis across vector, raster, and tabular data. \textit{OpenEarthAgent}~\cite{shabbir2026openearthagent} further advances this direction by providing a unified framework for tool-augmented geospatial reasoning trained on structured analytical trajectories, aligning the model with verified multi-step interactions over satellite imagery, natural language queries, and downstream tools. Similarly, \textit{Earth-Agent}~\cite{feng2025earth} expands the scope of agentic reasoning by unifying RGB, spectral, and geophysical modalities within a tool ecosystem designed for cross-modal and quantitative spatiotemporal analysis. Taken together, these systems illustrate a broader trend: progress in Earth science AI increasingly depends not only on stronger foundation models, but also on architectures that can operationalize scientific workflows through tool use. In this sense, tool-augmented agents serve as a bridge between general-purpose language intelligence and the procedural, data-intensive nature of Earth science research.

\subsubsection{Collaborative and Hierarchical Multi-Agent Architectures}

While single-agent systems are effective for many tasks, complex Earth science problems often require multiple forms of expertise that are difficult to robustly encode within a single agent. This has motivated the emergence of collaborative multi-agent architectures, in which different agents specialize in complementary roles such as planning, data retrieval, coding, visualization, and verification. Such modularization reflects the structure of real scientific workflows and improves scalability, controllability, and interpretability.

A representative example is \textit{ClimateAgent}~\cite{shan2026climateagents}, which introduces a role-specialized topology for complex climate data science workflows. Its \textit{Orchestrate-Agent} manages the global process and maintains context, the \textit{Plan-Agent} decomposes user queries into executable subtasks, \textit{Data-Agents} dynamically inspect APIs and generate robust data acquisition scripts, and \textit{Coding-Agents} execute and iteratively debug Python code to produce final analyses and visualizations. This explicit separation of responsibilities enables the system to tackle long-horizon workflows that would otherwise exceed the reliability of a monolithic agent. Other frameworks adopt similar principles in different forms. \textit{GeoColab}~\cite{wu2025geocolab} mimics a human geospatial development team by assigning agents the roles of Product Manager, Algorithm Engineer, and Programmer, thereby supporting collaborative geospatial code generation through structured interaction. \textit{PANGAEA-GPT}~\cite{pantiukhin2025accelerating} employs a hierarchical Supervisor--Worker architecture with strict data-type-aware routing and sandboxed deterministic code execution, enabling autonomous error diagnosis and recovery in geoscientific data discovery tasks. More broadly, these systems suggest that hierarchical and role-based decomposition is becoming a key architectural principle for Earth science agents. Collaborative agent networks are particularly promising for Earth science applications that span multiple interacting subsystems. Such cross-agent collaboration is well aligned with the inherently coupled nature of the Earth system, where meaningful scientific understanding often emerges only by integrating processes across the atmosphere, hydrosphere, biosphere, and human systems.

\subsubsection{Toward AI Earth Scientists}

The rise of agentic and multi-agent frameworks has also contributed to a broader conceptual shift toward \emph{AI Earth Scientists}: autonomous or human-in-the-loop scientific agents capable of integrating heterogeneous environmental and socioeconomic data, building predictive and explanatory models, generating hypotheses, and proposing scientifically grounded solutions. Under this view, the goal is no longer limited to automating isolated subtasks, but to support end-to-end scientific discovery and decision-making.

Recent systems begin to move in this direction. \textit{EarthLink}~\cite{guo2025earthlink}, for example, frames the agent as an interactive scientific copilot for climate research, integrating planning, code execution, data analysis, and physical reasoning into a unified workflow. Beyond static orchestration, such systems introduce self-evolving mechanisms, allowing the agent to improve through iterative feedback, debugging, and experience accumulation. This represents an important departure from conventional tool-calling pipelines: the agent is not merely executing predefined routines, but progressively refining how it conducts scientific analysis. Overall, these developments point toward a future in which Earth science AI is organized as an ecosystem of reasoning, tool use, collaboration, and adaptation. Rather than relying on a single all-purpose model, the field is increasingly moving toward modular, executable, and potentially self-improving scientific agents capable of operating across data modalities, software environments, and scales of Earth system inquiry.

\subsection{Future Directions: Embodied Earth Intelligence}

While recent agentic frameworks already move Earth science AI toward the vision of AI Earth Scientists, most existing systems still operate primarily in digital environments: they retrieve data, invoke tools, execute code, and reason over pre-collected observations. An important next step is to extend these capabilities toward \emph{embodied} and \emph{spatially grounded} intelligence, where agents can actively acquire data, interact with physical environments, and adapt their behavior during real-world scientific missions.

This direction is particularly relevant for Earth science because many critical tasks depend not only on passive analysis of archived observations, but also on adaptive sensing and autonomous exploration in challenging environments. Examples include underwater monitoring, disaster response, ecological surveying, and urban environmental inspection, where agents may need to decide where to observe, how to navigate, and which measurements to prioritize under uncertainty. In such settings, scientific reasoning must be tightly coupled with perception, memory, planning, and action. Recent embodied AI studies provide an early glimpse of this possibility. \textit{OceanGym}~\cite{xue2025oceangym}, for example, introduces a benchmark for underwater embodied agents operating in conditions characterized by low visibility, dynamic currents, and long-horizon mission requirements, highlighting the difficulty of robust perception and sequential decision-making in marine environments. In the broader spatial intelligence literature, systems such as \textit{EmbodiedScan}~\cite{wang2024embodiedscan}, \textit{Scene-LLM}~\cite{fu2024scene}, and \textit{SpatialBot}~\cite{cai2025spatialbot} demonstrate how multimodal agents can build grounded scene understanding, navigate complex spaces, and support action-oriented reasoning. Although these efforts are not yet tailored to Earth science, they suggest a natural extension of current Earth AI research: from tool-augmented software agents to embodied scientific agents capable of autonomous sensing, exploration, and data collection. Looking ahead, integrating foundation-model reasoning, agentic workflow orchestration, and embodied spatial intelligence may open a path toward more complete AI Earth Scientists. Such agents would not only analyze existing geoscientific data, but also actively seek new evidence, coordinate measurements across platforms, and close the loop between scientific hypothesis, observation, and action.

%% file: draft/042_challenges.tex
\section{Challenges and Outlook}
\label{sec:hallu}

Despite rapid progress, unified foundation models for Earth science still face substantial barriers before they can serve as reliable, general-purpose systems for scientific understanding, decision support, and autonomous discovery. These challenges arise not only from the scale and heterogeneity of Earth data but also from the need to ensure physical consistency, operational security, and effective interaction with complex scientific workflows.

\subsection{Key Challenges}

\subsubsection{Heterogeneous Data, Representation, and Earth Embeddings}

Earth science data are inherently heterogeneous, spanning satellite imagery, gridded reanalysis fields, in-situ observations, scientific text, and increasingly socioeconomic data. Constructing unified foundation models over such inputs remains difficult due to differing spatial and temporal resolutions, noise characteristics, and semantics. To overcome this, a primary challenge is to construct a unified \emph{Earth Embedding}---a compact yet semantically rich representation space learned across all Earth spheres. Such embeddings would allow researchers to exploit pre-learned representations for diverse downstream applications without repeatedly processing raw, high-resolution Earth observation (EO) data. Achieving this may also require cross-disciplinary inspiration, such as integrating state-space models for long spatiotemporal sequences or retrieval-augmented generation (RAG) to dynamically fetch relevant geospatial context.

\subsubsection{Scientific Reliability, Continual Updating, and Trustworthiness}

Unlike many general AI applications, Earth science models must remain scientifically reliable under evolving climates and high-stakes usage scenarios. Static pretraining quickly becomes outdated; thus, robust \emph{continual learning} frameworks are required to seamlessly integrate data from new satellite missions or upgraded sensor arrays without catastrophic forgetting. Furthermore, Earth AI systems must address critical security and privacy concerns. Because high-resolution multispectral imagery can expose sensitive human infrastructure, \emph{machine unlearning} mechanisms must be developed to selectively "forget" specific data subsets to comply with regulations. Simultaneously, integrating \emph{adversarial defenses} is crucial to protect models from subtly manipulated sensor inputs that could trigger disastrous miscalculations in extreme-event forecasting.

\subsubsection{Scalability, Efficiency, and Sustainable Adaptation}

Training and deploying large Earth science foundation models is computationally expensive because Earth data are high-dimensional and continuously growing. High-resolution spatial grids and multimodal fusion impose major burdens on storage, training, and inference. At the same time, the environmental cost of large-scale AI training is itself a concern in climate-related applications. Future progress will depend on energy-efficient adaptation strategies, including model quantization, parameter pruning, and interpretable knowledge distillation. These techniques are essential not only for reducing the carbon footprint of Earth AI but also for enabling real-time inference on edge devices, such as remote sensor networks and field-deployed monitoring stations.

\subsubsection{From Foundation Models to Agentic and Embodied Earth Intelligence}

A final challenge lies in moving beyond static prediction models toward systems that can reason, act, and collaborate in realistic scientific settings. The emerging Agent4Science paradigm addresses this by transitioning toward long-running, autonomous agents capable of executing multi-step workflows. By acting as persistent orchestrators equipped with domain-specific \emph{skills} (e.g., querying climate databases, executing numerical simulations, or optimizing spatial parameters), FMs can tackle complex professional tasks. Looking further ahead, extending these capabilities to embodied settings---such as autonomous environmental drones, underwater exploration, disaster response, and urban sensing---will require integrating spatial intelligence, active perception, and adaptive decision-making with Earth science reasoning. This shift from passive modeling to active, skill-equipped scientific agency will define the next stage of Earth AI.

\subsection{Outlook}

Looking ahead, the most promising direction is not simply to scale Earth science foundation models further, but to make them more \emph{integrated}, \emph{trustworthy}, and \emph{actionable}. First, the development of universal Earth Embeddings will democratize access to big Earth data, enabling highly efficient, cross-sphere applications. Second, Earth AI must become more reliable through provenance-aware continual updating, calibrated uncertainty estimation, and robust defenses against adversarial and privacy risks, particularly when projecting climate extremes in ungauged regions. 

In the longer term, the ultimate frontier is the convergence of unified Earth science foundation models, skill-based agentic reasoning, and embodied spatial intelligence. Such systems will not only analyze existing observations but also autonomously retrieve data, coordinate specialized scientific tools, plan sensing strategies, and interact with physical environments through robotic platforms. Realizing this vision will require close collaboration across geoscience, remote sensing, machine learning, and robotics. If successful, the field will progress from task-specific Earth AI toward a new generation of autonomous AI Earth Scientists capable of supporting end-to-end scientific discovery and planetary-scale stewardship.

%% file: draft/070_conclusion.tex
\section{Conclusion}
\label{sec:conclusion}



This survey has reviewed the emerging landscape of unified foundation models for Earth science and highlighted a clear transition from task-specific perception to systems capable of multimodal understanding, scientific reasoning, and autonomous agentic exploration. Collectively, these advances indicate that Earth AI is evolving from isolated predictive tools into general-purpose frameworks for holistic analysis and scientific discovery. Looking forward, addressing key challenges in data integration, scientific reliability, and real-world interaction will be essential to realizing the vision of AI Earth Scientists. We hope this survey provides a structured roadmap for future research and promotes deeper cross-disciplinary synergy between the AI and Earth science communities, thereby accelerating the development of intelligent, trustworthy, and actionable Earth system modeling.